\documentclass[sigplan, 10pt]{acmart}
\AtBeginDocument{%
  \providecommand\BibTeX{{%
    \normalfont B\kern-0.5em{\scshape i\kern-0.25em b}\kern-0.8em\TeX}}}

\copyrightyear{2025}
\acmYear{2025}
\setcopyright{acmlicensed}
\acmConference[EuroSys '25]{Twentieth European Conference on Computer Systems}{March 30-April 3, 2025}{Rotterdam, Netherlands}
\acmBooktitle{Twentieth European Conference on Computer Systems (EuroSys '25), March 30-April 3, 2025, Rotterdam, Netherlands}
\acmDOI{10.1145/3689031.3717461}
\acmISBN{979-8-4007-1196-1/25/03}

\usepackage{microtype}
\usepackage{graphicx}
\usepackage{booktabs} % for professional tables
\usepackage{tikz}
\usepackage{amsmath}
\usepackage{multirow}
\usepackage{fontawesome}
\usepackage{subcaption}
\usepackage{xspace}
\usepackage{booktabs}
\usepackage{comment}
\usepackage{float}
\usepackage{caption}
\usepackage{lipsum}
\usepackage[]{hyperref}
\usepackage{comment}
\usepackage{xspace}
\usepackage{wasysym}
\usepackage{float}
\usepackage{stfloats}
\usepackage{placeins}
\usepackage{enumitem}
\usepackage{xcolor}
\usepackage{booktabs}
\PassOptionsToPackage{table}{xcolor}
\PassOptionsToPackage{x11names}{xcolor}
\usepackage{xcolor}
\usepackage{colortbl}
\usepackage{makecell}
\usepackage{nicematrix}
\usepackage[ruled,linesnumbered]{algorithm2e}
\usepackage[export]{adjustbox}
\usepackage{float}
\usepackage[noend]{algpseudocode}

\usepackage{newtxmath}
\usepackage{pifont} % for cross mark

\definecolor{darkergreen}{RGB}{0,100,0} % Darker green
\definecolor{darkerred}{RGB}{139,0,0} % Darker red
\definecolor{lightergreen}{RGB}{34,139,34} % Lighter green, still subdued
\definecolor{lighterred}{RGB}{205,92,92} % Lighter red, still subdued
\usepackage{cleveref}
\crefformat{section}{\S#2#1#3} % see manual of cleveref, section 8.2.1
\crefformat{subsection}{\S#2#1#3}
\crefformat{subsubsection}{\S#2#1#3}

\usepackage{makecell}

\usepackage{xcolor} % Required for custom colors
\usepackage{listings} % Required for listings
\usepackage{pifont}% http://ctan.org/pkg/pifont
\definecolor{dmlgreen}    {RGB}{51,  160,  44}
\definecolor{dmlblue}     {RGB}{31,  120, 180}
\definecolor{dmlred}      {RGB}{202,   0,  32}
\definecolor{brown}       {RGB}{139,  69,  19}

\definecolor{deepblue}{rgb}{0,0,0.5}
\definecolor{deepred}{rgb}{0.6,0,0}
\definecolor{deepgreen}{rgb}{0,0.5,0}
\definecolor{mauve}{rgb}{0.58,0,0.82}
\definecolor{light-gray}{gray}{0.96}
\definecolor{aliceblue}{rgb}{0.94, 0.97, 1.0}
\definecolor{OliveGreen}{rgb}{0.33, 0.42, 0.18}

\usepackage[utf8]{inputenc}
\DeclareFixedFont{\ttb}{T1}{txtt}{bx}{n}{8} % for bold
\DeclareFixedFont{\ttm}{T1}{txtt}{m}{n}{8}  % for normal

\usepackage{xargs}                      % Use more than one optional parameter in a new commands
\usepackage[colorinlistoftodos,prependcaption,textsize=small]{todonotes}
\newcommandx{\note}[2][1=]{\todo[linecolor=OliveGreen,backgroundcolor=OliveGreen!25,bordercolor=OliveGreen,#1]{#2}}
\newcommandx{\unsure}[2][1=]{\todo[linecolor=red,backgroundcolor=red!25,bordercolor=red,#1]{#2}}
\newcommandx{\improve}[2][1=]{\todo[linecolor=orange,backgroundcolor=orange!25,bordercolor=orange,#1]{#2}}
\newcommandx{\change}[2][1=]{\todo[linecolor=Plum,backgroundcolor=Plum!25,bordercolor=Plum,#1]{#2}}
\newcommandx{\fix}[2][1=]{\todo[linecolor=blue,backgroundcolor=blue!25,bordercolor=blue,#1]{#2}}

\usepackage{titlesec}
\lstnewenvironment{env_python}[2]
{
\lstset{
  language=Python,
  backgroundcolor=\color{light-gray},
  otherkeywords={self},        
  emph={to,tile,reuse_at,reorder,unroll,split,vectorize,quantize},          
  emphstyle=\footnotesize\color{dmlred},  
  belowcaptionskip=0.7\baselineskip,
  aboveskip=-2mm,
  belowskip=5mm,
  showstringspaces=false,
  columns=flexible,
  escapechar=@,
  basicstyle={\linespread{0.99}\fontencoding{T1}\footnotesize\fontfamily{lmtt}\fontseries{m}\selectfont},
  numbers={left},
  xleftmargin={2em},%
  framesep=5pt, % margin top and down
  framexleftmargin={1em},
  numberstyle=\tiny\color{gray},
  keywordstyle=\color{blue},
  commentstyle=\color{deepgreen},
  stringstyle=\color{mauve},
  breaklines=true,
  breakatwhitespace=true,
  tabsize=3,
  frame=tb, 
  caption= #1, 
  label= #2
}%lstset
  \vspace{\baselineskip}
}%lstnewenvironment
{}

\newcommand{\lstbg}[3][0pt]{{\fboxsep#1\colorbox{#2}{\strut #3}}}
\lstdefinelanguage{diff}{
  backgroundcolor=\color{aliceblue},           
  emph={},          
  emphstyle=\small\color{dmlred},  
  belowcaptionskip=0.7\baselineskip,
  aboveskip=0mm,
  belowskip=3mm,
  showstringspaces=false,
  columns=flexible,
  basicstyle={\linespread{1.1}\fontencoding{T1}\scriptsize\fontfamily{lmtt}\fontseries{m}\selectfont},
  numbers={left},
  xleftmargin={2em},%
  breaklines=true,
  breakatwhitespace=true,
  tabsize=3,
  morecomment=[f][\lstbg{red!20}]-,
  morecomment=[f][\lstbg{green!20}]+,
  morecomment=[f][\textit]{@@},
}

\makeatletter
\let\old@lstKV@SwitchCases\lstKV@SwitchCases
\def\lstKV@SwitchCases#1#2#3{}
\makeatother
\usepackage{lstlinebgrd}
\makeatletter
\let\lstKV@SwitchCases\old@lstKV@SwitchCases

\lst@Key{numbers}{none}{%
    \def\lst@PlaceNumber{\lst@linebgrd}%
    \lstKV@SwitchCases{#1}%
    {none:\\%
     left:\def\lst@PlaceNumber{\llap{\normalfont
                \lst@numberstyle{\thelstnumber}\kern\lst@numbersep}\lst@linebgrd}\\%
     right:\def\lst@PlaceNumber{\rlap{\normalfont
                \kern\linewidth \kern\lst@numbersep
                \lst@numberstyle{\thelstnumber}}\lst@linebgrd}%
    }{\PackageError{Listings}{Numbers #1 unknown}\@ehc}}
\makeatother
\newcommand{\Name}{Mist\xspace}
\newcommand{\myparagraph}[1]{\noindent\textbf{{#1}.\;}}
\newcommand{\cmark}{\ding{51}} % pifont checkmark
\newcommand{\xmark}{\ding{55}} % pifont cross

\newcommand{\filledtwo}{\ding{183}\xspace}
\newcommand{\filledthree}{\ding{184}\xspace}

\definecolor{customcolor}{HTML}{f9ae78}
\definecolor{customcolor2}{HTML}{3d5c6f}

\usepackage{xcolor}
\usepackage{listings}
\lstdefinestyle{mystyle}{
    basicstyle=\ttfamily\small,      % Small font
    numbers=left,                    % Line numbers on the left
      numberstyle=\color{gray},   % Small gray numbers
}


\begin{document}

\title{Mist: Efficient Distributed Training of Large Language Models via Memory-Parallelism Co-Optimization}

\author{Zhanda Zhu}
\affiliation{%
  \institution{University of Toronto, Vector Institute, CentML}
  \country{}
  }
\email{zhanda.zhu@mail.utoronto.ca}

\author{Christina Giannoula}
\affiliation{%
  \institution{University of Toronto, Vector Institute, CentML}
  \country{}
  }
\email{christina.giann@gmail.com}

\author{Muralidhar Andoorveedu}
\affiliation{%
  \institution{CentML}
  \country{ }
  }
\email{murali@centml.ai}

\author{Qidong Su}
\affiliation{%
  \institution{University of Toronto, Vector Institute, CentML}
  \country{}
  }
\email{qdsu@cs.toronto.edu}

\author{Karttikeya Mangalam}
\affiliation{%
  \institution{SigIQ.ai}
  \country{}
  }
\email{mangalam@sigiq.ai}

\author{Bojian Zheng}
\affiliation{%
  \institution{University of Toronto, Vector Institute, CentML}
  \country{}
  }
\email{bojian@cs.toronto.edu}

\author{Gennady Pekhimenko}
\affiliation{%
  \institution{University of Toronto, Vector Institute, CentML}
  \country{}
  }
\email{pekhimenko@cs.toronto.edu}

\renewcommand{\shorttitle}{Mist: Efficient Distributed Training of Large Language Models via Memory-Parallelism ...}
\renewcommand{\shortauthors}{Z. Zhu, C. Giannoula, M. Andoorveedu, Q. Su, K. Mangalam, B. Zhang, G. Pekhimenko}

\begin{abstract}
Various parallelism, such as data, tensor, and pipeline parallelism, along with memory optimizations like activation checkpointing, redundancy elimination, and offloading, have been proposed to accelerate distributed training for Large Language Models.
To find the best combination of these techniques, automatic distributed training systems are proposed.
However, existing systems only tune a subset of optimizations, due to the lack of overlap awareness, inability to navigate the vast search space, and ignoring the inter-microbatch imbalance, leading to sub-optimal performance.
To address these shortcomings, we propose \Name, a memory, overlap, and imbalance-aware automatic distributed training system that comprehensively co-optimizes all memory footprint reduction techniques alongside parallelism.
\Name is based on three key ideas:
(1) fine-grained overlap-centric scheduling, orchestrating optimizations in an overlapped manner,
(2) symbolic-based performance analysis that predicts runtime and memory usage using symbolic expressions for fast tuning, and 
(3) imbalance-aware hierarchical tuning, decoupling the process into an inter-stage imbalance and overlap aware Mixed Integer Linear Programming problem and an intra-stage Dual-Objective Constrained Optimization problem, and connecting them through Pareto frontier sampling.
Our evaluation results show that \Name achieves an average of 1.28$\times$ (up to 1.73$\times$) and 1.27$\times$ (up to 2.04$\times$) speedup compared to state-of-the-art manual system Megatron-LM and state-of-the-art automatic system Aceso, respectively.

\end{abstract}

\maketitle

\section{Introduction}

% 1.1 Importantce of Optimizing Distributed Training for LLMs
Large Language Models (LLMs) have gained high interest and show remarkable capabilities in various fields like question answering, summarization, problem solving, and more~\cite{chatgpt, llama3, claude}.
However, their significantly increased sizes and dataset requirements have escalated computational and memory demands.
For instance, training LLaMa-3.1-405B~\cite{llama3.1} uses a cumulative 30.84M GPU hours of computation on NVIDIA H100 GPUs~\cite{llama3.1-training-cost}. While most companies and researchers cannot afford to pre-train such LLMs, continuous pre-training or supervised fine-tuning still costs over 2000 NVIDIA H100 GPU hours per 1B tokens~\cite{llama3.1-training-cost}.
Therefore, efficient distributed training techniques have been proposed~\cite{megatron-1, deepspeed, pytorch-distributed, pytorch-fsdp, wang2019tofu, tarnawski2021piper, flexflow, unger2022unity, li2022amp, athlur2022varuna, lai2023merak, zheng2022alpa, miao2023galvatron, liu2024aceso, chen2023slapo, sun2024adapipe, lin2024nnscaler, atc24-yuan-hybrid-offloading} to improve system performance during training, since even minor reductions in the training time are translated to significant financial and environmental benefits~\cite{strubell2019energy, llama2-training-cost, llama3-training-cost}.

% 1.2 Techniques used in large scale distributed training
Prior works~\cite{megatron-1,megatron-2,deepspeed,gpipe,pipedream,zheng2022alpa,flexflow} propose various parallelization techniques for  distributed training.
Data Parallelism (DP)~\cite{pytorch-distributed, tensorflow} splits input data across devices, with each device processing a portion of the data, while maintaining a full copy of the LLM model.
Tensor Parallelism (TP)~\cite{megatron-1,megatron-2,megatron-3} partitions the parameters of each layer across devices, however introducing inter-device communication over activations to ensure computation correctness.
Pipeline Parallelism (PP)~\cite{gpipe, pipedream, pipetransformer, terapipe} divides the model into stages, however a large number of stages may lead to performance inefficiencies, e.g., pipeline bubbles where devices are idle during training.
We henceforth refer to the number of partitions in DP, TP and PP methods as DP size, TP size, and PP size, respectively.
Gradient accumulation techniques are usually applied, dividing a global batch into multiple microbatches, to reduce the memory pressure of each microbatch and facilitate pipeline parallelism~\cite{megatron-1}.
To alleviate memory pressure in devices, memory footprint reduction techniques~\cite{sublinear-memory-cost-checkpointing, echo, checkmate, dtr,zero, pytorch-fsdp,rhu2016vdnn, peng2020capuchin, zero-offload, zero-infinity, tempo, feng2023mobius, guo2024gmlake} have also been proposed. 
Activation checkpointing (CKPT)~\cite{sublinear-memory-cost-checkpointing, echo, checkmate, dtr} reduces memory footprint during training by recomputing activations during backpropagation. 
ZeRO~\cite{zero, pytorch-fsdp} eliminates model states redundancy by partitioning the optimizer states, gradients, or weights across devices. Higher ZeRO levels partition more model states, thereby providing larger memory footprint savings, however increasing the inter-device communication.
Offloading~\cite{rhu2016vdnn, peng2020capuchin, zero-offload, zero-infinity, feng2023mobius, guo2024gmlake} temporarily transfers unused tensors from the GPU to the CPU, freeing GPU memory however increasing the data transfer costs.
These memory optimizations often require overlapping data transfers with GPU computation to reduce performance overhead ~\cite{peng2020capuchin, zero-offload, pytorch-fsdp}.

In this work, we observe that memory footprint reduction techniques, although they have been primarily designed to alleviate memory pressure, they can significantly enhance performance, since they assist in balancing trade-offs between runtime overhead and memory footprint reduction.
For instance, applying offloading optimization can free up some memory in GPU devices, which can then be leveraged to reduce the TP or PP size, thereby reducing communication overheads or pipeline bubbles during training.
Generally, exploiting memory footprint reduction techniques to release some memory footprint in GPU devices can be leveraged to:
1) reduce the TP size, thus mitigating communication overheads; 
2) reduce the PP size, thus eliminating pipeline bubbles; 
% 3) decrease the number of layers that need to be recomputed;  
% 4) decrease ZeRO level, reducing communication overhead;
% 5) lower offloading ratios, reducing CPU-GPU communication overhead; 
and (3) increase the batch size, improving kernel efficiency.
Conversely, applying less aggressively memory footprint reduction optimizations results in higher GPU memory usage, which increases the partitioning across devices, thus incurring higher performance overheads related to parallelism.
Therefore, additional GPU memory can be gained by applying more aggressive memory footprint reduction techniques, which come with some added overhead. This memory can then be used to reduce the overhead of other optimizations. As long as the benefit from reducing the overhead outweighs the additional cost incurred by the memory footprint reduction techniques, overall training efficiency improves. See detailed motivational examples in Section~\ref{sec:the-need-for-comprehensive-co-optimization}.

% 1.3 Existing Distributed Training Techniques
Overall, distributed training constitutes an optimization problem that can be formulated as choosing the best combination of all available techniques (both parallelism and memory footprint reduction techniques) to maximize training efficiency, while keeping the memory usage lower than the available hardware memory capacity.
Manual distributed training methods such as Megatron-LM~\cite{megatron-1} and DeepSpeed~\cite{deepspeed}, among others~\cite{zero-offload, zero-infinity, pytorch-fsdp}, are developed to provide some of the above optimizations.
However, these manual methods require users to specify configurations, i.e., the combination of parallelism and memory footprint reduction techniques, for optimal performance. This can be quite challenging even for experienced users and takes lots of engineering efforts~\cite{zheng2022alpa}.
Moreover, as model sizes and device counts increase, tuning complexity increases exponentially~\cite{zheng2022alpa}.
To address this issue, automatic distributed training systems have been proposed~\cite{wang2019tofu, tarnawski2021piper, flexflow, unger2022unity, lai2023merak, zheng2022alpa, miao2023galvatron, liu2024aceso, chen2023slapo, sun2024adapipe, lin2024nnscaler, atc24-yuan-hybrid-offloading}.
Given LLM model and GPU hardware, they construct the search space of various configurations of the their supported optimizations and automatically find optimal combination of them.

We extensively examine distributed training frameworks and identify a key shortcoming: 
they fail to comprehensively co-optimize memory footprint reduction techniques alongside parallelism, since they only focus on a subset of the available search space.
Specifically, these systems are still inefficient in optimizing training performance, because they
(i) either tune parallelism configuration with a fixed pre-defined memory optimization~\cite{flexflow, zheng2022alpa, miao2023galvatron}, 
(ii) support only one specific optimization like activation checkpointing~\cite{liu2024aceso, sun2024adapipe, lai2023merak}, 
or (iii) make strong assumptions, such as applying the same memory footprint strategy across all pipeline stages~\cite{atc24-yuan-hybrid-offloading}. 
These constraints lead to a reduced search space and sub-optimal performance, as demonstrated in Section~\ref{sec:the-need-for-comprehensive-co-optimization}.

% 1.4 Limitations of Existing Distributed Training Approaches
We analyze how prior works tune the training configurations and find that co-optimizing all available memory footprint reduction techniques and parallelism is challenging in these prior existing systems, because they suffer from three limitations.
First, existing automatic systems do \emph{not} overlap communication with computation beyond the basic gradient all-reduce, thus missing important opportunities for training efficiency.
This can cause severe performance degradation (See Figure~\ref{fig:eval-2-main-without-FlashAttn}), which becomes even more severe when all memory optimizations are involved.
% (1) fail to co-optimize
Second, they are not able to efficiently explore the exploded search space when co-tuning all optimizations. When more memory optimization techniques are incorporated, the search space increases significantly, and existing systems fail to find the best configuration in such a huge search space.
% (3) fail to be inter-microbatch imbalance aware
Third, they use the averaged microbatch time to model the pipeline parallelism performance, implicitly assuming uniform microbatch execution time within a pipeline stage.
However, we find that this is not the case, since first and last microbatches incur higher communication costs (especially when ZeRO and offloading are involved), as we explain in Section~\ref{sec:shortcomings}.

% 1.5 Proposed ideas
To tackle the aforementioned limitations, we propose \Name, a memory, overlap, and imbalance-aware automatic distributed training system that co-optimizes memory footprint reduction techniques with parallelism.
\Name consists of three key ideas:
(1) \textbf{fine-grained overlap-centric scheduling}, which carefully orchestrates the implementation of both memory footprint reduction techniques and parallelism to maximize the computation-communication overlapping;
(2) \textbf{symbolic-based efficient performance analysis}, which enables fast exploration of the exploded search space of various configurations by efficiently predicting runtime and memory usage via symbolic expressions and batched value substitutions;
and (3) \textbf{imbalance-aware hierarchical tuning}, which takes into account the microbatch variability and overlap opportunities in pipeline parallelism, decouples the optimization process into an inter-stage Mixed Integer Linear Programming (MILP) problem and an intra-stage Dual-Objective Constrained Optimization problem, and connects them via Pareto frontier sampling. This third key idea addresses both the search space explosion and the microbatch imbalance in pipeline parallelism.

We extensively evaluate \Name on a wide variety of LLMs, including GPT-3~\cite{brown2020gpt3}, LlaMa~\cite{llama2}, and Falcon~\cite{almazrouei2023falcon}, across diverse training configurations, i.e., varying the global batch size, model size, and using different kernel implementations such as FlashAttention~\cite{dao2022flashattention}) and hardware setups (up to 32 NVIDIA L4~\cite{nvidia-l4} and 32 NVIDIA A100 GPUs~\cite{nvidia-a100}), and demonstrate that \Name significantly outperforms prior state-of-the-art works~\cite{megatron-1, deepspeed, liu2024aceso}.
Our evaluation results show that \Name achieves an average of 1.28$\times$ (up to 1.73$\times$) and 1.27$\times$ (up to 2.04$\times$) speedup compared to the state-of-the-art manual implementation Megatron-LM and the state-of-the-art automatic system Aceso, respectively, across different GPU, model, and training configurations.

% 1.6 Key Evaluation and Contributions
To sum up, we make the following contributions:

\begin{itemize}[topsep=1.5pt, partopsep=0pt, leftmargin=*, itemsep=1pt]
    \item We identify the need of comprehensively co-optimizing memory footprint reduction techniques alongside parallelism and propose \Name, a highly efficient easy-to-use automatic distributed training framework for LLMs.
    \item We propose and implement a symbolic analysis system that generates symbolic expressions for workload characteristics to quickly explore the exploded search space.
    We design a scheduling method that maximizes computation-communication overlap by carefully coordinating memory optimization and parallelism techniques.
    We introduce a tuning method that decouples the optimization process into two stages and connects them through Pareto frontier sampling, addressing microbatch variability and leveraging overlap opportunities in pipeline parallelism.
    \item We evaluate \Name using various large-scale LLMs in both NVLink systems (NVIDIA A100 GPUs~\cite{nvidia-a100}) and PCIe systems (NVIDIA L4 GPUs~\cite{nvidia-l4}) and compare it to multiple strong baselines. \Name significantly outperforms prior works, up to $2.04\times$, under various training configurations.
\end{itemize}
\section{Background}
\label{sec:background-and-motivation}

\begin{figure}[t]
    \centering
    \includegraphics[width=0.9\linewidth]{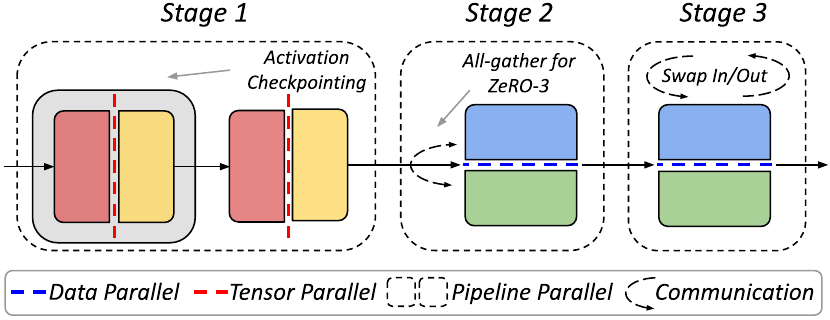}
    \caption{An illustration of optimization configurations.}
    \label{fig:background-optimization-illustration}
\end{figure}

LLMs~\cite{brown2020gpt3, llama} require excessive computation and memory, leading to significant costs and energy consumption~\cite{llama3.1-training-cost}.
Consequently, distributed training, i.e., scaling hardware and splitting the model and/or the input data, is the typical solution to train LLMs~\cite{megatron-1, deepspeed}.
To efficiently conduct distributed training, various parallelism techniques~\cite{pytorch-distributed, megatron-1, gpipe} and GPU memory footprint reduction methods~\cite{sublinear-memory-cost-checkpointing, peng2020capuchin, feng2023mobius, zero} have been developed.
As shown in Figure~\ref{fig:background-optimization-illustration}, different parallelism and memory optimizations can be applied in combination.

\subsection{Parallelism in Distributed Training}
\label{subsec:parallelism-in-distributed-training}

\myparagraph{Data Parallelism}
To scale training, \emph{Data Parallelism (DP)}~\cite{tensorflow, pytorch-distributed} distributes input data across GPUs, with each GPU processing its data independently using a model replica. 
It involves only an all-reduce of gradients per iteration, but requires the entire model to fit within each GPU's memory.

\myparagraph{Tensor Parallelism}
\emph{Tensor Parallelism (TP)}~\cite{megatron-1, megatron-2} partitions the parameters in each layer and conducts all-reduce over activations in the forward pass and gradients in the backward pass to maintain computation correctness. 

\myparagraph{Pipeline Parallelism}
\emph{Pipeline Parallelism (PP)}~\cite{gpipe, pipedream, dapple, memory-efficient-pp, terapipe, torchgpipe} partitions the model into stages, using p2p communication between stages to pass data through the pipeline.
Although it only involves small communication overhead to transfer intermediate tensors, the dependency between stages introduces pipeline bubbles, which causes efficiency to suffer as a result of the device idle time.

\begin{table}[t]
\footnotesize
\centering
% \captionsetup{skip=4pt, belowskip=-9.5pt}
\caption{Comparison of distributed training systems. 
P, G, O, and A under offloading denote parameter, gradient, optimizer states, activation offloading, respectively.
Circle for optimizations represents functionality support and granularity.
Circle for tuning represents whether the system can tune all optimizations it supports.
}
 \setlength{\tabcolsep}{1.85pt}
 \begin{tabular}{l c c c c c c c c c c}
 \toprule
    % Optimization & \makecell{DP} & \makecell{SDP} & \makecell{TP} & \makecell{PP} & \makecell{Act. Ckpt.} & \makecell{Offloading} \\ 
    & \multicolumn{3}{c}{\makecell[c]{Parallelism}} & \multicolumn{4}{c}{\makecell[c]{Offloading}} & \multirow{2}{*}{\makecell{ZeRO-\\2/3}} & \multirow{2}{*}{\makecell{Auto-Tuning\\Capability}} \\
    & DP & TP & PP & P & G & O & A & \\
 \midrule
    Megatron-LM~\cite{megatron-1}      & \cmark & \cmark & \cmark & \Circle     & \Circle     & \Circle     & \Circle     & \xmark & \Circle     \\
    DeepSpeed~\cite{deepspeed}         & \cmark     & \cmark & \cmark & \Circle     & \Circle     & \Circle     & \Circle     & \cmark & \Circle     \\
    ZeRO-Offload~\cite{zero-offload}   & \cmark     & \cmark & \xmark & \Circle     & \Circle     & \LEFTcircle & \Circle     & \cmark & \Circle     \\
    ZeRO-Infinity~\cite{zero-infinity} & \cmark     & \cmark & \xmark & \LEFTcircle & \LEFTcircle & \LEFTcircle & \LEFTcircle & \cmark & \Circle     \\
 \midrule
    Alpa~\cite{zheng2022alpa}          & \cmark     & \cmark & \cmark & \Circle     & \Circle     & \Circle     & \Circle     & \cmark & \LEFTcircle \\
    Slapo~\cite{chen2023slapo}         & \cmark     & \cmark & \cmark & \Circle     & \Circle     & \Circle     & \Circle     & \cmark & \LEFTcircle \\
    AdaPipe~\cite{sun2024adapipe}      & \cmark     & \cmark & \cmark & \Circle     & \Circle     & \Circle     & \Circle     & \xmark & \LEFTcircle \\
    Yuan et al.~\cite{atc24-yuan-hybrid-offloading} & \cmark & \cmark & \cmark & \Circle     & \Circle     & \Circle     & \LEFTcircle     & \xmark & \LEFTcircle     \\
    Aceso~\cite{liu2024aceso}          & \cmark & \cmark & \cmark & \Circle     & \Circle     & \Circle     & \Circle     & \xmark & \CIRCLE     \\
 \midrule
    Mist                               & \cmark     & \cmark & \cmark & \CIRCLE     & \CIRCLE     & \CIRCLE     & \CIRCLE     & \cmark & \CIRCLE     \\
 \bottomrule
 \end{tabular}
\label{tab:comparison}
\end{table}

\subsection{GPU Memory Footprint Reduction Techniques}
\label{subsec:gpu-memory-footprint}

\myparagraph{Activation Checkpointing}
\emph{Activation checkpointing (CKPT)} (also known as \emph{recomputation})~\cite{sublinear-memory-cost-checkpointing, dtr, checkmate, echo, tempo} discards certain activations in the forward pass, while stashing others. 
Later in the backward pass, the discarded activations are recomputed from the stashed activations, and are then used for gradient computation. 
This method reduces the memory needed for the saved activations, at the cost of recomputing discarded activations in the backward pass.

\myparagraph{Redundancy Optimization} \textit{Zero Redundancy Optimizer}
(\textit{ZeRO}) reduces the memory usage by eliminating redundant copies of optimizer states, gradients, and model weights across data-parallel devices~\cite{zero, pytorch-fsdp}.
ZeRO operates in three modes: ZeRO-1 (shards optimizer only), ZeRO-2 (shards optimizer and gradients), and ZeRO-3 (shards optimizer, gradients, and weights). While ZeRO-1 introduces no additional communication, ZeRO-2/3 incur communication overhead due to all-gathering and reduce-scattering operations.

\myparagraph{Offloading}
\textit{Offloading}~\cite{rhu2016vdnn, peng2020capuchin, wang2018superneurons, zero-infinity, feng2023mobius, guo2024gmlake} (also known as \emph{swapping}) involves transferring model states or activations between the GPU devices and the host CPU. This technique helps manage GPU memory constraints by temporarily offloading data, allowing the GPU to accommodate other tasks.
The efficiency of swapping significantly depends on overlapping, which allows memory transfers to be hidden outside the critical path.

\section{Limitation of Existing Systems}

\subsection{The Need for Comprehensive Co-Optimization}
\label{sec:the-need-for-comprehensive-co-optimization}

Distributed training optimization problem involves finding the optimal configuration of parallelism strategies and memory reduction methods to maximize performance, given specific hardware, model, and global batch size.
Current distributed training systems, however, lack the ability to comprehensively co-optimize memory footprint optimizations alongside parallelism~\cite{zheng2022alpa, liu2024aceso, atc24-yuan-hybrid-offloading}.
As detailed in Table 1, manual methods, such as ZeRO-Infinity~\cite{zero-infinity}, support a broader range of memory optimizations but only allow coarse-grained configuration (enabling or disabling offloading) and lack automatic tuning. Automatic methods, such as Alpa~\cite{zheng2022alpa}, AdaPipe~\cite{sun2024adapipe}, and Aceso~\cite{chen2023slapo}, either support fewer optimizations or can only tune a subset of the optimizations they provide. This limited search space leads to sub-optimal configurations and reduced performance.
% Instead, they focus on a subset of the search space, as detailed in Table~\ref{tab:comparison}, resulting in sub-optimal configurations and performance.

\begin{figure}[t]
    \centering
    \includegraphics[width=0.975\linewidth,left]{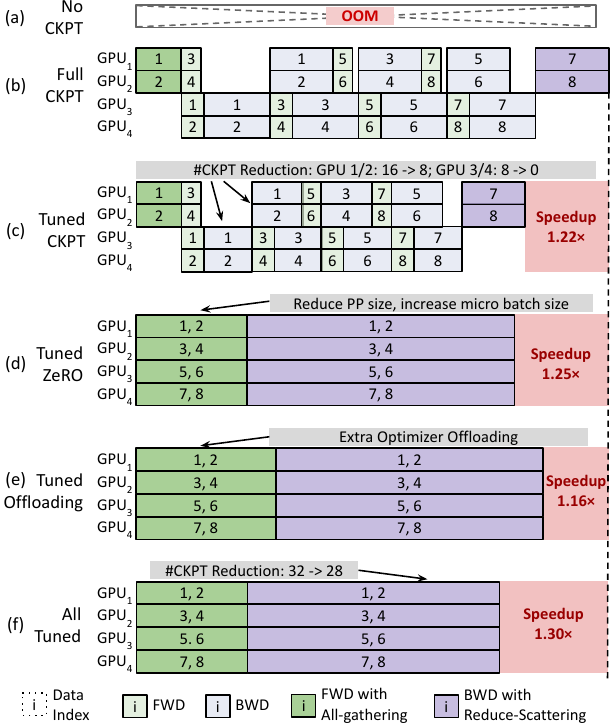}
    \caption{Motivational examples of tuning parallelism with memory optimizations for GPT-3-2.7B on 4 NVIDIA L4 GPUs with $Seq=4096, B_{global}=8$. Parallelism is always tuned.}
    \label{fig:single-memory-opt}
\end{figure}

To illustrate these trade-offs of different optimizations and the impact of co-optimization, we present a motivational example of training GPT-3-2.7B on four NVIDIA L4 GPUs with a global batch size of 8.
We manually enumerate all $DP$, $TP$, $PP$, and micro batch size $b$ configurations, and when co-tuning memory optimizations, we also enumerate their combinations with the parallelism configurations.

As shown in Figure~\ref{fig:single-memory-opt}(a), without memory optimization, all parallelism plans result in out-of-memory (OOM) errors.
In Figure~\ref{fig:single-memory-opt}(b), applying full CKPT (all layers being recomputed, as in Megatron-LM~\cite{megatron-1} and Alpa~\cite{zheng2022alpa}) reduces memory usage by recomputing activations, avoiding OOM.
The best parallelism strategy found is $DP$=2, $PP$=2, $b$=1.
In Figure~\ref{fig:single-memory-opt}(c), if activation checkpointing is tuned (as in Aceso~\cite{liu2024aceso} and Adapipe~\cite{sun2024adapipe}), the number of recomputed layers is reduced from 16 to 8 on the first two GPUs, and from 16 to 0 on the other two, reducing recomputation.
During tuning, although another strategy ($DP$=1, $PP$=4, $b$=1) fully eliminates recomputation by using the extra memory from the increased PP size, the added pipeline bubbles outweigh the benefits of reduced recomputation, causing it to under-perform compared to $PP$=2.
In Figure~\ref{fig:single-memory-opt}(d), tuning ZeRO (as in DeepSpeed~\cite{deepspeed}) enables $DP$=4, $PP$=1, $b$=2 with ZeRO-2, preventing OOM by sharding gradients. 
Similarly, in Figure~\ref{fig:single-memory-opt}(e), tuning offloading enables the same parallelism with an optimizer offloading ratio of 0.325, avoiding OOM.
In both cases, reduced pipeline bubbles and improved kernel efficiency (from the increased batch size) outweigh the memory optimization overhead, increasing training efficiency.
These examples show that tuning each memory optimization with parallelism improves training performance, achieving speedups of $1.22\times$, $1.25\times$ and $1.16\times$ for CKPT, ZeRO, and offloading tuning, respective, compared to the full CKPT strategy.

Building upon these findings, we co-optimize all memory optimizations with parallelism and identify an even better strategy: $DP$=4, $PP$=1, $b$=2 with ZeRO-2 and adjusted activation checkpointing (recomputed layers reduced from 32 to 28), which reduces pipeline bubbles (compared to activation checkpointing tuning only) and recomputation (compared to ZeRO tuning only), leading to a 1.30$\times$ speedup while maintaining memory savings.

\begin{figure}[t]
    \centering
    \small

    % \begin{minipage}[c]{0.325\linewidth}
    %     \centering
    %     \includegraphics[width=\linewidth]{figs/motivation-1-1.pdf}
    % \end{minipage}
    % \hfill
    % \begin{minipage}[c]{0.325\linewidth}
    %     \centering
    %     \includegraphics[width=\linewidth]{figs/motivation-1-2.pdf}
    % \end{minipage}
    % \hfill
    % \begin{minipage}[c]{0.325\linewidth}
    %     \centering
    %     \includegraphics[width=\linewidth]{figs/motivation-1-3.pdf}
    % \end{minipage}\vspace{2pt}
    % \textbf{(a)} Tuning Parallelism Only\\
    
    % \vspace{10pt}  % Add vertical space between rows
    
    % Second row
   % \begin{minipage}[c]{0.325\linewidth}
   %      \centering
   %      \includegraphics[width=0.95\linewidth]{figs/motivation-2-1.pdf}
   %  \end{minipage}
   %  \hfill
   %  \begin{minipage}[c]{0.325\linewidth}
   %      \centering
   %      \includegraphics[width=\linewidth]{figs/motivation-2-2.pdf}
   %  \end{minipage}
   %  \hfill
   %  \begin{minipage}[c]{0.325\linewidth}
   %      \centering
   %      \includegraphics[width=\linewidth]{figs/motivation-2-3.pdf}
   %  \end{minipage}\vspace{2pt}
   %  \textbf{(a)} Tuning Parallelism and Checkpointing\\

    % \vspace{10pt}  % Add vertical space between rows
    
    % Third row
    \begin{minipage}[c]{0.325\linewidth}
        \centering
        \includegraphics[width=\linewidth]{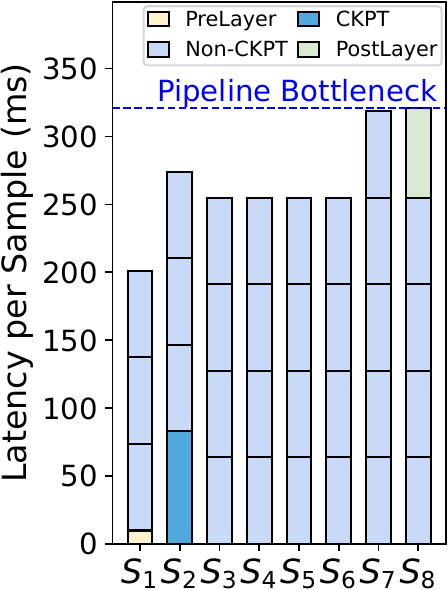}
        {\footnotesize \textbf{(a)} Runtime of tuning parallelism with CKPT}
    \end{minipage}
    \hfill
    \begin{minipage}[c]{0.325\linewidth}
        \centering
        \includegraphics[width=\linewidth]{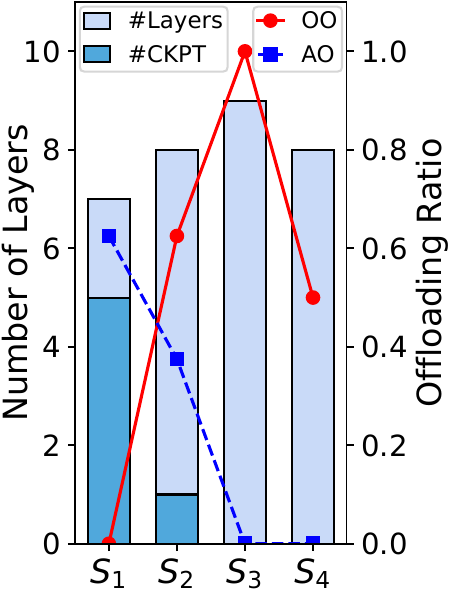}
        {\footnotesize \textbf{(b)} Configurations of comprehensive tuning}
    \end{minipage}
    \hfill
    \begin{minipage}[c]{0.325\linewidth}
        \centering
        \includegraphics[width=\linewidth]{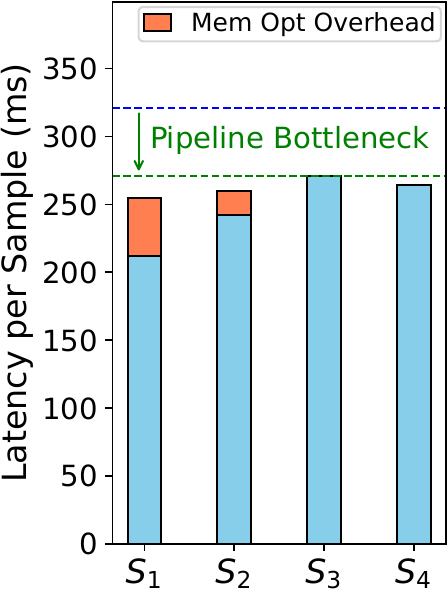}
        {\footnotesize \textbf{(c)} Runtime of comprehensive tuning}
    \end{minipage}    
    % \hfill
    % \begin{minipage}[c]{0.24\linewidth}
    %     \centering
    %     \includegraphics[width=\linewidth]{figs/motivation-3-3.pdf}
    % \end{minipage}\vspace{2pt}
    % \textbf{(b)} Tuning all Memory Optimizations with Parallelism\\[2pt]
        
    % Caption for the entire figure
    \caption{Motivational example of showing the speedup source of comprehensive co-optimization for GPT-3-7B on 8 NVIDIA L4 GPUs with $Seq = 2048$, $B_{global}=512$.}
    \label{fig:motivational-example}
\end{figure}

To further demonstrate the benefits of comprehensive co-optimization, we consider an example of training GPT-3-7B on eight NVIDIA L4 GPUs with a global batch size of 512. 
When only activation checkpointing is tuned, the best parallelism strategy identified is $DP$=1, $PP$=8, $b$=1, which causes severe pipeline imbalance and hardware idling, as shown in Figure~\ref{fig:motivational-example}(a). 
However, by comprehensively co-optimizing all techniques, we find a better strategy: $DP$=2, $PP$=4, $b$=2, with adjusted activation checkpointing and optimized offloading ratios, detailed in Figure~\ref{fig:motivational-example}(b), where $OO$ and $AO$ stand for optimizer and activation offloading, respectively.
This configuration uses offloading to gain GPU memory, which is then used to reduce PP size from 8 to 4 and eliminate recomputation for the last two stages.
As shown in Figure~\ref{fig:motivational-example}(c), co-optimization reduces pipeline stages and device idle time, improving overall performance despite some offloading overhead, as the optimizer offloading overhead is amortized over multiple micro-batches and activation offloading can overlap with computation.
Comprehensive co-optimization yields a $1.22\times$ speedup over tuning only parallelism and a $1.11\times$ speedup over tuning parallelism with activation checkpointing, demonstrating significant performance gains.

However, existing systems lack support for comprehensive co-optimization.
For instance, Aceso~\cite{liu2024aceso} does not support ZeRO or offloading, Slapo~\cite{chen2023slapo} only tunes activation checkpointing within a fixed parallelism plan, and AdaPipe~\cite{sun2024adapipe} focuses solely on pipeline parallelism and activation checkpointing. 
It limits their ability to fully leverage the trade-offs between memory reduction and runtime overhead, leading to suboptimal performance. 

\subsection{Why Existing Auto Systems Fail to Co-Optimize?}
\label{sec:shortcomings}

We summarize the key shortcomings of existing systems in achieving comprehensive co-optimization:

\begin{figure}[t]
    \centering
    \includegraphics[width=0.99\linewidth]{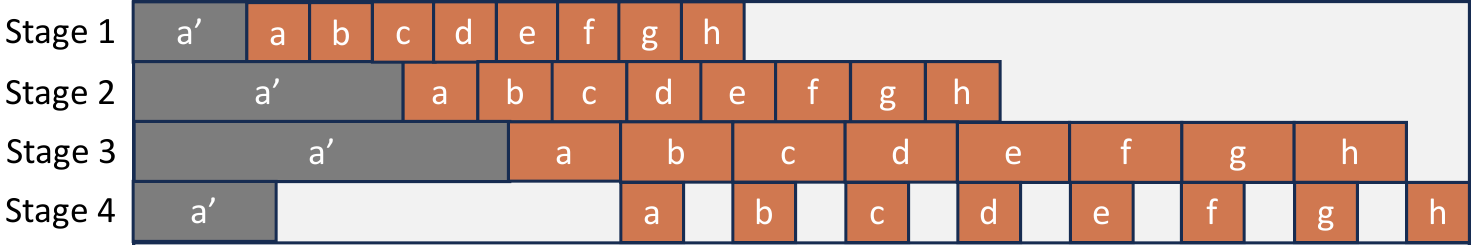}
    % \rule{\linewidth}{2.5cm}
    \caption{Illustration of pipeline parallelism overlap opportunity and inter-microbatch imbalance. $\text{a}'$ is the extra communication happened in the first microbatch.}
    \label{fig:challenge-pipeline-overlap}
\end{figure}

\myparagraph{Shortcoming \#1: Lack of overlap awareness}
Existing automatic distributed training methods fail to account for computation-communication overlap beyond basic gradient synchronization overlap.
This results in significant performance degradation, as seen in our experiments where Aceso underperforms manual implementation Megatron-LM (with overlap) in 6 out of 10 cases despite a larger search space (See Figure~\ref{fig:eval-2-main-without-FlashAttn}).
Moreover, techniques like ZeRO and offloading add extra communication overheads, requiring overlap with computation or pipeline bubbles to maintain efficiency~\cite{zero,zero-infinity}.
In Figure~\ref{fig:motivational-example}(b), Stage 2 shows a 13\% overhead if activation offloading is not overlapped, and for Stage 3, offloading optimizer states for a 7B model with a $PP=4$ takes 7 seconds, resulting in a 40\% overhead with a batch size of 64.
Additionally, computation-communication interference results in inaccurate performance predictions.
For instance, we observe a 7.7\% performance degradation for the linear layer in attention module of the motivational example when it is executed concurrently with all-reduce operations, which becomes worse when CPU-GPU communication is also involved.
Ignoring overlap leads to mis-estimating optimization configurations and results in sub-optimal strategies.

\begin{figure}[t]
    \centering
    % \rule{0.5\linewidth}{2.5cm}
    \includegraphics[width=0.95\linewidth]{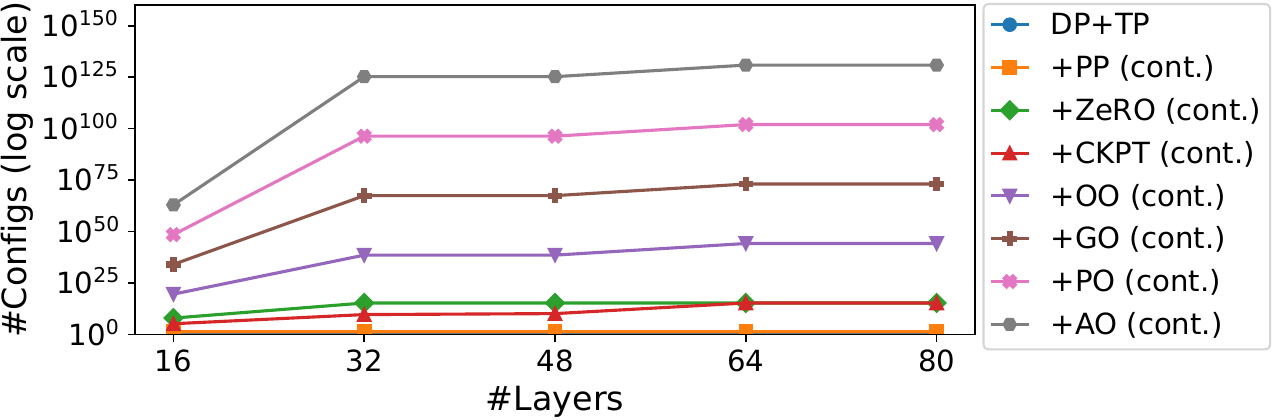}
    \caption{Growth in the number of configurations within the search space as each optimization is incrementally added.}
    \label{fig:search-space}
\end{figure}

\myparagraph{Shortcoming \#2: Unable to navigate the exploded search space}
Co-optimizing memory footprint reduction techniques with parallelism significantly expands the search space, making it difficult to efficiently find the best combination.
For example, Alpa takes over 40 hours to find the best parallelization strategy for GPT-3-39B on 64 GPUs~\cite{zheng2022alpa}.
As depicted in Figure~\ref{fig:search-space}, simultaneously tuning parallelism and memory optimizations further dramatically increases the search space and complexity.
Even after applying search space pruning methods, such as inter and intra-stage tuning decoupling, the search space remains significantly larger than what existing performance predictors can efficiently handle~\cite{santhanam2021distir, geoffrey2021habitat, hu2022dpro, distsim, duan2023proteus}.
For example, Proteus, a fast simulation-based tool that supports the prediction of performance in parallelization and recomputation, requires around 6 seconds to simulate \textit{one} optimization configuration for GPT-2 on 32 GPUs~\cite{duan2023proteus}. 
Despite its speed, this kind of tool is still impractical for effectively exploring the vast search space presented by our problem.

\myparagraph{Shortcoming \#3: Inaccurate performance prediction due to the lack of inter-microbatch imbalance awareness}
Existing automatic distributed training systems often suffer from inaccurate performance predictions when new memory optimizations are involved due to their inability to account for inter-microbatch variability.
Automatic parallelism planners~\cite{zheng2022alpa, liu2024aceso} typically assume uniform microbatch execution times within a pipeline stage. However, the first and last microbatches take longer due to extra operations like parameter all-gathering, gradient reduce-scattering, and optimizer offloading, as shown in Figure~\ref{fig:challenge-pipeline-overlap}.
In the motivational example of tuning GPT-3-7B, simply averaging microbatch times leads to performance prediction error of up to 21.55\%, depending on the number of microbatches, which may cause up to a twofold performance slowdown.
Our evaluation, as shown in Figure~\ref{fig:eval-3-ablation}, ignoring inter-microbatch imbalance leads to about a 9\% slowdown compared to optimal solutions.
These inaccuracies undermine the effectiveness of tuning process, leading to the selection of sub-optimal strategies.

\subsection{Why Simple Heuristics Can Not Address it?}

Why manual frameworks with simple heuristics cannot address co-optimization?
Manual frameworks struggle to deliver optimal performance in comprehensive co-optimization scenarios, because the best performance can only be achieved with fine-grained configuration for each stage, including layer assignments, DP and TP sizes, recomputed layers, and offloading ratios for each type of model states.
The vast and exponentially large search space makes manual exploration impractical, leading users to rely on simple heuristics.
For example, a recent study~\cite{atc24-yuan-hybrid-offloading} uses a heuristic that applies uniform checkpointed layers and activation offloading ratios across all pipeline stages to reduce tuning search space. 
However, pipeline parallelism inherently exhibits memory and computation imbalances, making uniform strategies sub-optimal.
As shown in Figure~\ref{fig:motivational-example}(b), heterogeneous optimizations per stage are selected.
In our motivational examples, uniform heuristics results in 26\% and 20\% performance degradation for the 2.7B and 7B models, respectively, compared to the optimal strategies achieved through comprehensive co-optimization.
As workloads scale, the complexity of tuning grows, making fully automated systems increasingly essential to efficiently handle the expanded search space and deliver optimal performance~\cite{zheng2022alpa, liu2024aceso, sun2024adapipe}.

% P4: Our goal
\paragraph{Our Goal} \textit{ is to develop a fully automated distributed training optimization system for LLMs that addresses all the shortcomings above and comprehensively co-optimize memory footprint reduction techniques alongside parallelism.}

\section{Mist: Overview and Key Ideas}

To accelerate distributed training, we introduce \Name, a memory, overlap, and imbalance aware automatic distributed training system that comprehensively co-optimizes memory footprint reduction techniques with parallelism. 
Overall, \Name proposes three key ideas:

\myparagraph{1. Fine-Grained Overlap-Centric Scheduling and Interference Modeling}
\Name proposes an overlap-centric scheduling approach that carefully orchestrates parallelism and memory optimizations to maximize the overlap of computation and communication.
By optimizing the order and granularity of these techniques, \Name mitigates memory optimization overhead while maintaining manageable tuning complexity.
Additionally, data-driven interference modeling accurately predicts performance when computation and communication kernels run concurrently.
This approach addresses Shortcoming \#1, the lack of overlap awareness.

\myparagraph{2. Symbolic-Based Efficient Performance Prediction}
Building upon the scheduling strategy, \Name introduces a symbolic analysis system to significantly enhance the performance prediction efficiency.
Unlike traditional methods that require repeated simulations for each optimization configuration~\cite{zheng2022alpa, duan2023proteus, liu2024aceso}, \Name symbolizes the model and optimizations, requiring only a single simulation pass to predict runtime and memory usage in the form of symbolic expressions. 
Subsequent predictions are simplified to value substitutions in these expressions, dramatically reducing redundant simulation and allowing for rapid batched evaluation of multiple configurations.
This approach effectively mitigates the Shortcoming \#2 search space explosion.

\myparagraph{3. Imbalance-Aware Hierarchical Tuning via Pareto Frontier Sampling} 
Finally, \Name proposes an imbalance-aware hierarchical auto-tuner that decouples tuning into intra-pipeline-stage and inter-pipeline-stage processes, while still considering the microbatch imbalances and overlap opportunities in PP.
To find the best intra-stage configurations, our intra-stage tuning uses a dual-objective approach to balance time between imbalanced and stable microbatches. Inter-stage tuning formulates an MILP problem to determine the best pipeline partitioning using data points sampled from the Pareto frontier of intra-stage tuning.
This approach preserves a pruned search space from hierarchical tuning (helping to solve Shortcoming \#2) while addressing the lack of inter-microbatch imbalance awareness (Shortcoming \#3).

\begin{figure}[t]
    \centering
    \includegraphics[width=0.9\columnwidth]{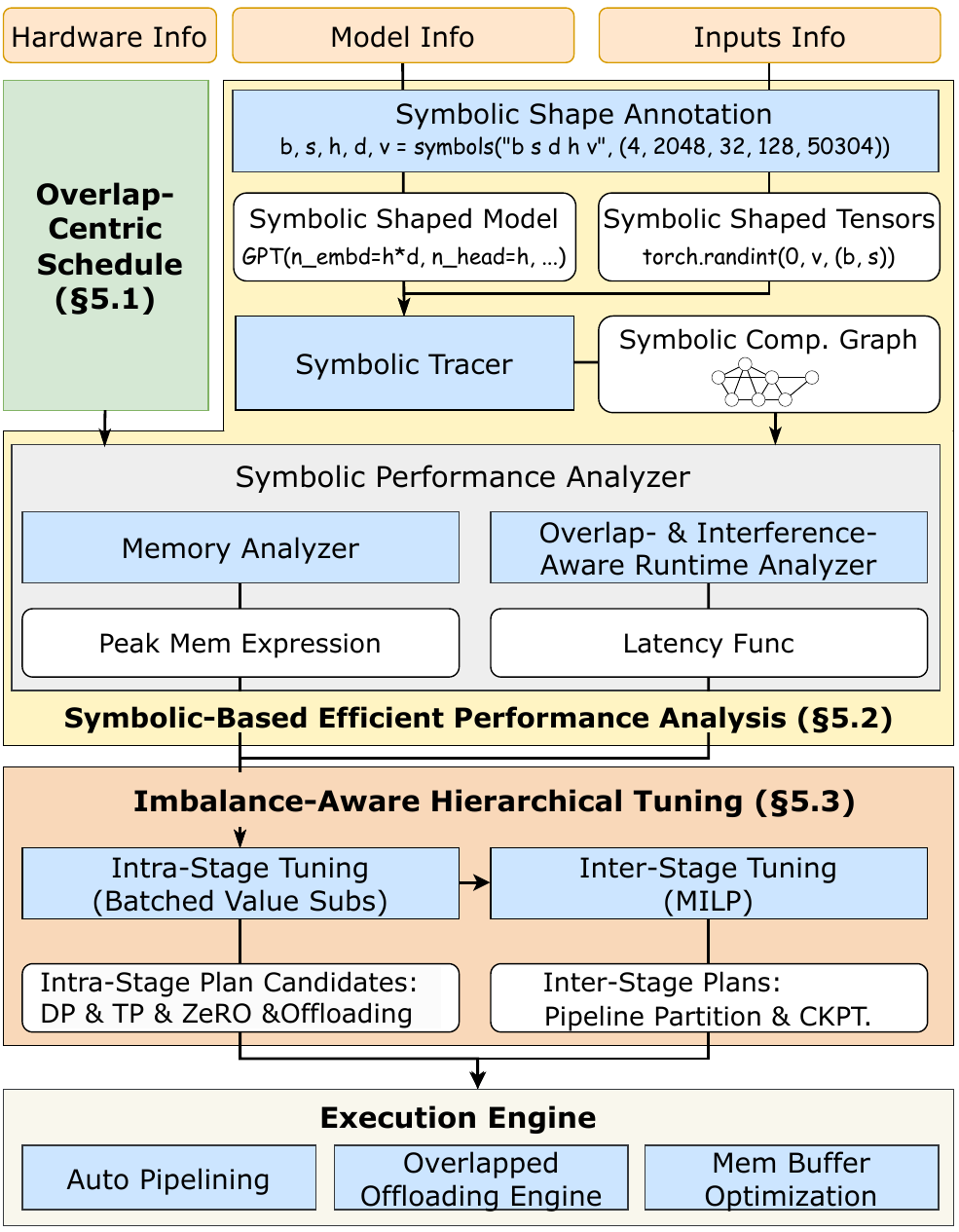}
    \caption{High-level System Overview of \Name.}
    \label{fig:overview}
\end{figure}

\myparagraph{System Overview}
Figure~\ref{fig:overview} presents the high-level overview of \Name. 
The model and its input data are annotated with symbolic shapes and traced to generate a symbolic shaped computational graph.
% The model and its input data are annotated with symbolic shapes and traced to generate a symbolic shaped computational graph.
This graph, on top of the Overlap-Centric Scheduling, is analyzed by our symbolic performance analyzer to derive the peak memory expression and runtime function whose inputs are optimization-related symbols.
During intra-stage tuning, \Name evaluates these symbolic expressions with specific optimization values in batches. This evaluation identifies a Pareto-optimal set of parallelism and memory optimization plans for each potential inter-stage candidate, utilizing our Symbolic-Based Efficient Performance Prediction strategy to efficiently navigate the extensive search space.
Subsequently, the inter-stage tuning phase formulates a MILP problem using stable microbatch times ($T_{stable}$) and their deviations ($T_{delta}$) sampled from intra-stage tuning results. This determines the optimal pipeline partitioning and combination of ($T_{stable}$, $T_{delta}$), addressing both inter-microbatch imbalances and inter-stage imbalances in pipeline parallelism.
Once the optimal tuning plans are identified, \Name employs an orchestrated execution engine to execute them, including automatic pipeline transformation, overlapped offloading communication, and memory buffer optimizations.
\section{Mist: Design Details}

\subsection{Fine-Grained Overlap-Centric Scheduling}
\label{sec:fine-grained-overlap-centric-scheduling}

\begin{figure*}[!t]
    \centering
    \includegraphics[width=1.0\textwidth]{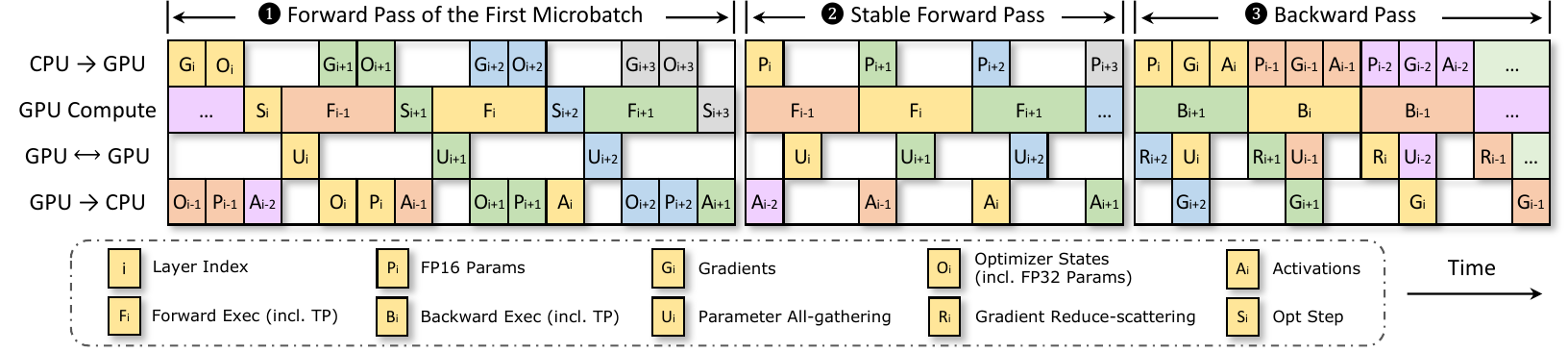}
    \caption{Overlap Schedule Template of \Name}
    \label{fig:schedule}
\end{figure*}

\Name coordinates memory optimizations and parallelism with two primary objectives: balancing optimization effectiveness with manageable tuning complexity and maximizing overlap opportunities to reduce runtime overhead.

\begin{table}[t]
% \small
\footnotesize
\centering
\caption{Optimization variables in the schedule template.}
 \begin{tabular}{l l l} 
 \toprule
    \textbf{Name} &  \textbf{Value Type} & \textbf{Meaning} \\ 
 \midrule
    $G$  & Integer  & Gradient accumulation steps \\
    $S$ & Integer  & Number of pipeline stages \\
 \midrule
    $L_i$  & Integer & Number of layers in stage $i$ \\
    $b_i$  & Integer & Micro batch size for stage $i$ \\
    $DP_i$ & Integer & DP size for stage $i$\\
    $TP_i$ & Integer & TP size for stage $i$ \\
 \midrule 
    $ZeRO_i$ & One-Hot [0-3] & ZeRO level for stage $i$ \\
    $CKPT_i$ & Integer & Number of recomputed layers for stage $i$ \\
    $WO_i$ & Float [0, 1] & Weight offloading ratio for stage $i$ \\
    $GO_i$ & Float [0, 1] & Gradient offloading ratio for stage $i$ \\
    $OO_i$ & Float [0, 1] & Opt states offloading ratio for stage $i$ \\
    $AO_i$ & Float [0, 1] & Activation offloading ratio for stage $i$ \\
 \bottomrule
 \end{tabular}
\label{tab:optim-schedule-explaination}
\end{table}

\myparagraph{Optimizations and Granularity}
\Name comprehensively supports various parallelism techniques such as DP, TP, and PP, alongside memory optimizations like fine-grained offloading, flexible activation checkpointing, and different levels of ZeRO optimization.
Based on the observation that all transformers are identical and share computational properties within a pipeline stage, \Name adopts stage-wise tuning granularity, meaning all layers within the same pipeline stage use the same parallelism and memory optimization configurations to balance optimization effectiveness and search space.
Specifically, for gradient accumulation steps $G$ and the number of pipeline stages $S$, the combination of ($L_i$, $b_i$, $DP_i$, $TP_i$, $ZeRO_i$, $CKPT_i$, $WO_i$, $GO_i$, $OO_i$, $AO_i$) defines the configuration for pipeline stage $i$ (see Table~\ref{tab:optim-schedule-explaination} for detailed explanations of the symbols). 
Notably, swapping strategies are represented as floating-point ratios, enabling fine-grained control of memory offloading and enhancing computation-communication overlap potential.

\myparagraph{Overlapped Schedule}
\label{sec: schedule-overlapped-schedule}
Building upon these optimization strategies, the overlapped schedule optimizes the execution order of memory optimizations and parallelism techniques to maximize hardware utilization while maintaining a low GPU memory footprint.
% Fwd and Bwd explanation
As depicted in Figure~\ref{fig:schedule}, computation, GPU-GPU communication, and CPU-GPU communication are overlapped.
As shown in \filledtwo,
during the forward pass, the computation of layer $k$ overlaps with the activation swapping out of layer $k-1$, and the swapping in and all-gathering of parameters for layer $k+1$.
Similarly, as shown in \filledthree, in the backward pass, the computation of layer $k$ overlaps with the gradient reduction and the swapping-out of the previous backward layer $k+1$, along with the swapping in of parameters, gradients and activations, and all-gathering of parameters for the next layer $k-1$.
% Conclusion
This overlap ensures that the computation in layer $k$ is not stalled by data movement or pre-fetching, leading to better hardware utilization.
% PP Overlap
Additionally, \Name supports inter-stage overlap by hiding communications that are independent of previous stages within pipeline bubbles, as shown in Figure~\ref{fig:pipeline-overlap-design-detail}.

\myparagraph{Optimizer Step Decoupling and Repositioning}
\label{sec: schedule-decouple-and-reposition-the-opt-step}
% Obersevations.
Furthermore, in scenarios involving ZeRO optimization and offloading, a monolithic optimizer step can lead to increased peak memory usage and redundant communication.
Specifically, to perform an optimizer step in the mixed-precision optimizer, the following tensors must be in the GPU device at the same time: FP16 parameters, FP16 gradients, FP32 optimizer states, and FP32 master parameters~\cite{zero}. 
% Observation 1
When offloading is applied, peak memory during the optimizer step may exceed that of the backward pass, as only partial layer states reside in GPU memory during the forward and backward computations, while a monolithic optimizer step requires all of them for the whole model. 
% Observation 2
Additionally, optimizer steps require rematerializing all states through offloading or all-gathering, which also occur during the forward and backward passes, introducing redundant communication.
% Solution
To address these issues, \Name decouples the optimizer step into multiple steps, repositioning each layer's optimizer step immediately before its first forward pass. Moreover, to eliminate synchronization between pipeline stages for \texttt{nan} and \texttt{inf} checks, the validate-and-update method from zero-bubble pipeline parallelism can be employed, delaying synchronization and reverting optimizer states if necessary~\cite{qi2023zero-bubble-pp}.

\subsection{Symbolic-Based Efficient Performance Analysis}
\label{sec:symbolic-analysis}

\begin{figure}[t]
    \centering
    \includegraphics[width=0.95\linewidth]{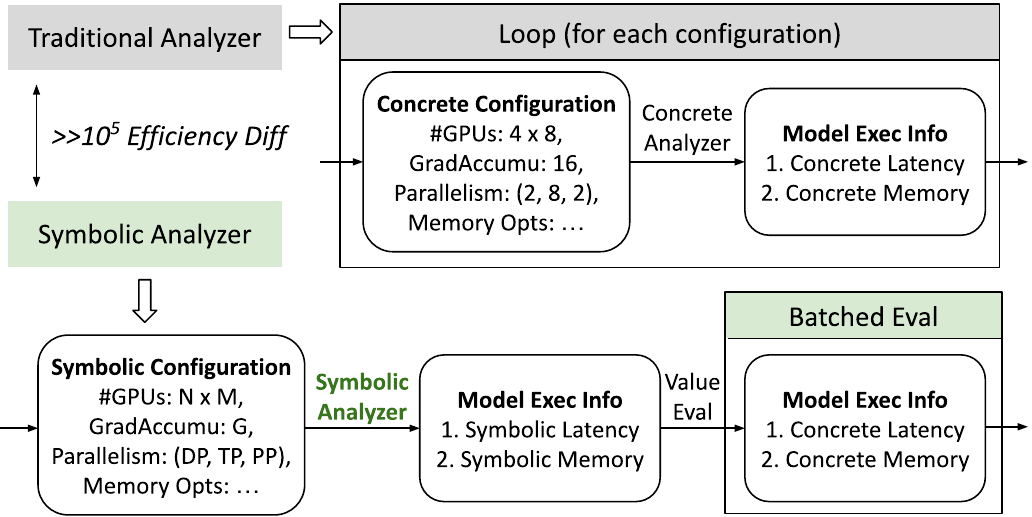}
    \caption{Comparison of the symbolic performance analyzer with the traditional analyzer.}
    \label{fig:symbolic-analysis}
\end{figure}

Performance modeling is crucial for optimizing distributed training as it enables efficient configuration exploration. Existing systems rely on simulation-based performance prediction, running a concrete simulation for each configuration to estimate computation, communication, and memory usage. 
As shown in Figure~\ref{fig:symbolic-analysis}, a traditional simulator initializes a GPT model with a concrete parallelism configuration of e.g., ($DP$=2, $TP$=8, $PP$=2) on 32 GPUs, applies memory optimizations, and simulates execution to measure performance and peak memory usage. Although each simulation is efficient, it still takes about 6 seconds per configuration~\cite{duan2023proteus}. This cost makes exhaustive search impractical for our combined search space. We now demonstrate how \Name's symbolic-based performance analysis helps to accelerate the performance prediction and thus facilitates the traversal over a huge search space.

\subsubsection{Symbolic Analysis System}
\Name overcomes these limitations by employing symbolic-based performance modeling.
Symbolic analysis refers to techniques used to analyze systems by reasoning about symbolic representations of data or computations rather than concrete values, used in compiler optimizations~\cite{cadar2008klee} and circuit designs~\cite{floberg2012symbolic-circuit-design}.
\Name proposes a symbolic analysis system for LLMs, supporting symbolic execution, tracing, and analysis.
As shown in Figure~\ref{fig:symbolic-analysis-code}, users define the model and inputs, simply replacing the concrete dimensions and optimizations with symbols.
Then all operations are executed with the information of symbolic shapes.
On top of it, \Name traces the computational graph and performs static analysis to derive symbolic expressions for execution time and memory usage. Instead of repeatedly simulating different configurations, \Name only performs a single symbolic simulation pass and later substitutes values into these expressions to quickly evaluate different configurations. This approach enables batched evaluation and compilation optimization, making performance prediction over $10^{5}\times$ faster than traditional analyzers.

\begin{figure}
\begin{lstlisting}[language=python,basicstyle=\ttfamily\scriptsize]
from mist import global_symbol_manager as gsm
# Define symbols
b, s, h, d, tp = gsm.symbols("b s h d tp", (4, 128, 12, 64, 8), integer=True, positive=True)
# Initialize the model configuration using symbolic parameters
config = GPT2Config(n_embd=h*d, n_head=h, tp=tp, ...)
# Construct the GPT-2 model with symbolic configuration
model = GPT2LMHeadModel(config)
# Create symbolic input tensors
input_ids = torch.randint(0, V, (b, s), dtype=torch.long)
# Execute the model with symbolic inputs
logits = model(input_ids).logits
\end{lstlisting}
\begin{lstlisting}[language=python,basicstyle=\ttfamily\scriptsize,numbers=none, xleftmargin=0pt, framexleftmargin=0pt]
>>> logits
symbolic_tensor((b, s, V), concrete_shape=(4, 128, 50257), ...)
\end{lstlisting}
\caption{Example of defining symbolic model configurations and inputs, followed by symbolic execution.}
\label{fig:symbolic-analysis-code}
\vspace{-5pt}
\end{figure}

For memory analysis, \Name uses liveness analysis on the symbolic computational graph. It tracks live tensors during execution and determines peak memory usage by identifying the maximum memory allocation at any point. 
To support pipeline parallelism, \Name performs intra-layer and inter-layer analysis: the intra-layer pass extracts memory statistics (e.g., layer states, saved activations, and intermediate tensors), while the inter-layer pass combines this data to generate stage-wise symbolic memory expressions, enabling efficient estimation across configurations.

For runtime analysis, direct symbolic representation is impractical due to the complex behavior of various GPU kernels. Instead, \Name profiles operator execution dynamically. Computation is estimated using a operator computation database, which benchmarks new operators or unseen input shapes on the current hardware and stores results for future use. Communication is modeled symbolically by dividing communicated bytes by the bandwidth, and overlap is managed via interference modeling, which we introduce below.

The design of our symbolic analysis system is far from trivial and addresses several significant challenges, making it both powerful and widely applicable.
First, large models must be run across multiple GPUs due to memory capacity issues, but direct analysis on multi-GPU setups is inefficient. 
We solve this by using the idea of fake tensors and meta devices, where tensor shapes are represented symbolically but not materialized physically, allowing analysis without needing actual hardware.
Second, backward pass memory analysis is difficult due to the absence of an explicit computational graph. 
We generate a fake backward graph using gradient function properties to track memory during backpropagation. 
Third, supporting custom kernels like FlashAttention~\cite{dao2022flashattention} and communication operations required custom symbolic representations, ensuring flexibility. 
Beyond optimization, our symbolic analysis system offers clear insights into workloads, making it easier to understand how specific parameters and optimizations affect performance, which can be valuable for both practical use and educational purposes.

\subsubsection{Interference Model}
\label{sec:interference-model}
% Challenge
To be overlap aware, we integrate an interference model within the symbolic analysis system.
Runtime prediction is much more challenging when overlap is involved, such as computation, NCCL (GPU $\leftrightarrow$ GPU communication), D2H (GPU$\rightarrow$CPU communication), and H2D (CPU$\rightarrow$GPU communication) running simultaneously.
\Name provides an interference model that predicts the impact of up to four different types of kernels running simultaneously. Instead of using machine learning models like XGBoost~\cite{chen2016xgboost}, which may overfit in this case, we develop a mathematical model with fewer parameters and clearer intuition. In this model, each possible combination of co-running kernels is assigned a set of slowdown factors that quantify the effect of the execution for each participant.

Algorithm~\ref{algo:interference} implements batched interference estimation, which iteratively applies slowdown factors to update execution times.
For each concurrency level ($n=$ 4 to 2 operations), it iterates through all $\binom{4}{n}$ combinations, retrieves predefined masks and factors, and invokes \texttt{Update}.
The \texttt{Update} function scales execution time by their respective slowdown factors, computes the scaled overlapping, and updates remaining execution times accordingly.
By progressively resolving interference through successive reductions, the algorithm eliminates concurrent operations until only a single component remains.
A data-driven approach is used to fit the model, where different shapes and combinations of concurrent kernels are sampled and benchmarked, and the resulting runtime data is used to train the slowdown factors.

\begin{algorithm}[t]
\footnotesize
\caption{Batched Interference Estimation}
\label{algo:interference}
\DontPrintSemicolon  % Don't print semicolons at the end of each line
\KwData{$C, G2G, C2G, G2C$, $params$}
\KwResult{Total latency vector $T$}

\SetKwFunction{FMain}{PredINTF}
\SetKwFunction{FUpdate}{Update}
\SetKwProg{Fn}{Function}{:}{}
\SetKwProg{Pn}{Function}{:}{\KwRet}

\Fn{\FMain{$C, G2G, C2G, G2C, params$}}{
    $X \gets [C, G2G, C2G, G2C]^T$ \tcp*{Stack features}
    $T \gets ZerosLike(C)$\tcp*{Initialize output}

    \For{$n = 4$ \textbf{downto} 2\tcp*{The num of concurrent ops}}{
        \For{$i = 0$ \textbf{to} $\binom{4}{n} - 1$\tcp*{Enumerate all combinations}}{
            \tcp{Index pre-defined masks and factors}
            $mask \gets IndexPredefinedMask(n, i)$ \;
            $factors \gets IndexFactors(params, n, i)$ \;
            \FUpdate{$X, T, mask, factors$}\;
        }
    }
    
    % \tcp{Add for only one active}
    $T \mathrel{{+}{=}} \text{sum}(X, \text{axis}=-1)$\tcp*{Sum remaining time}
    \KwRet $T$\;
}

\Pn{\FUpdate{$X, T, mask, factors$}}{
    % $\mathit{ids} \gets \{j \mid \left( X_j \neq 0 \right) \equiv mask\}$ \\
    $\mathit{ids} \gets \{j \mid \left( X_j \neq 0 \right) \text{ matches } mask\}$ \\
    \lIf{$\mathit{ids} = \emptyset$}{\Return}
    $\mathit{scaled} \gets X[\mathit{ids}] \times factors$ \\
    % $indices \gets \text{where}(\text{all}(X \neq 0 \text{ matches } mask, \text{axis}=-1))$\;
    % $scaled \gets X[indices] \times factors$\;
    $overlap \gets \min(scaled, \text{axis}=-1)$\;
    $X[\mathit{ids}] \gets (scaled - overlap) / factors$\;
    $T[\mathit{ids}] \mathrel{{+}{=}} overlap$\;
}
\end{algorithm}

\subsection{Imbalance-Aware Hierarchical Tuning via \\Pareto Frontier Sampling}
\label{sec: tuner}

Our tuning problem is defined as, given a model, a global batch size $B$, and a device mesh $(N, M)$, \Name's auto-tuner outputs the best training plan including the gradient accumulation steps $G$, layer partitions $PP$ for different pipeline stages, and combination of [$b$, $DP$, $TP$, $ZeRO$, $CKPT$, $WO$, $GO$, $OO$, $AO$] for each pipeline stage, as detailed in Section~\ref{sec:fine-grained-overlap-centric-scheduling}.

% General Methods
To efficiently find the best strategy in a huge search space, \Name adopts the idea of hierarchical tuning, decoupling the whole tuning process into intra-stage tuning and inter-stage tuning~\cite{zheng2022alpa}. 
Intra-stage tuning aims at finding the best optimization plans for all possible pipeline partitioning candidates, while inter-stage tuning is used to find the best stage partitioning and device assignment.
Compared to existing automatic parallelization methods, \Name offers two key improvements: imbalance and overlap awareness.

\begin{figure}[t]
    \centering
    \includegraphics[width=0.95\columnwidth]{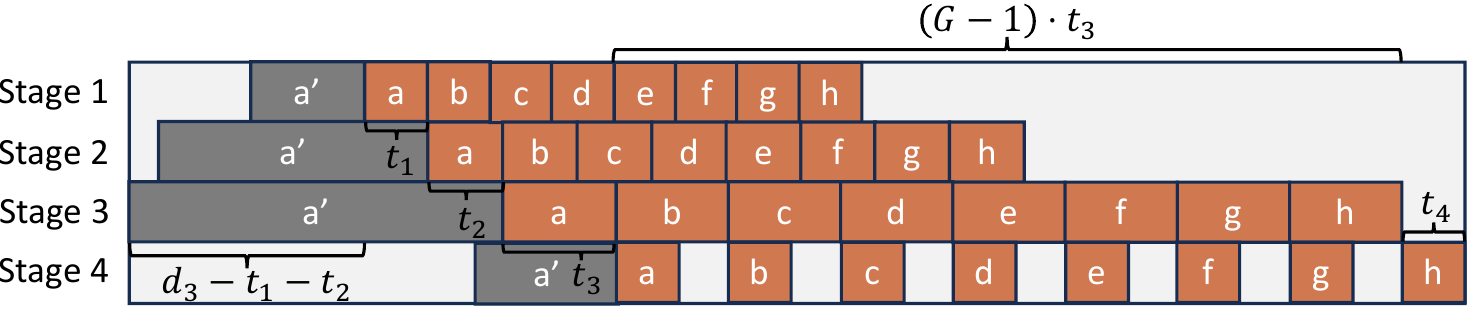}
    \caption{Illustration of runtime of a pipeline considering inter-microbatch imbalance. a' is the extra communication overhead only involved in the first and last microbatches.}
    \label{fig:pipeline-overlap-design-detail}
\end{figure}

\myparagraph{Inter-Stage Tuning}
Inter-stage tuning finds the best layer partition and device assignment, as well as the best number of layers being recomputed.
% Difference
% Newly added
Unlike previous methods that treat all microbatches as the same~\cite{zheng2022alpa,liu2024aceso}, \Name discovers that using either averaged runtime across different microbatches or simply applying the stable microbatch to tune leads to sub-optimal results.
% Why not good
% Newly added
The former approximation might lead to bottleneck drifting, and the latter one fails to consider the extra overhead of optimizations specifically for the first or last micro batch.
% Mist's solution
As Figure~\ref{fig:pipeline-overlap-design-detail} shows, \Name considers the inter-microbatch imbalance and proposes a new objective as 
\begin{equation}
    \min_{\substack{1 \leq i \leq S \\ l_i, ckpt_i, \\(n_i, m_i)}}
    % \min_{\{l_i, (n_i, m_i)\}_{i=1}^S}
    \left\{ 
        (G-1) \cdot \max_{1 \leq i \leq S} \left\{ t_{i} \right\} + \sum_{i=1}^{S} t_{i} + \max_{1 \leq i \leq S} \left( d_i - \sum_{1 \leq j < i} t_{i}  \right) 
    \right\}
    \label{equ:inter-op-obj-without-pareto}
\end{equation}
where $S$ means the number of stages, $l_i$ denotes the number of layers, $ckpt_i$ denotes the number of checkpointed layers, and $\left(n_i, m_i\right)$ denotes the device assignment, for stage $i$.
For simplicity, we define any microbatch that is neither first nor last as a stable microbatch.
$t_i$ means the \textit{stable} microbatch runtime of stage $i$, and $d_i$ means the runtime \textit{delta} of the first and last microbatches compared to $t_i$.
The first term ensures that \Name correctly identifies the pipeline bottleneck, while the second and third terms account for inter-stage and inter-microbatch imbalances, respectively.
The third term also considers the overlap opportunities of hiding communication independent of previous stages in the pipeline bubbles.

Objective~\ref{equ:inter-op-obj-without-pareto} can be solved given ($t_i$, $d_i$) according to $l_i$, $c_i$, and $(n_i, m_i)$.
However, $t_i$ and $d_i$ are correlated within a stage.
For instance, if optimizer offloading is applied aggressively, the runtime of the first microbatch significantly increases and the runtime of the stable microbatches reduces because of the less intensive memory pressure.
This suggests that ($t_i$, $d_i$) form pairs along a Pareto frontier within the stage. 
Thus, we transform the decision variables and obtain:
% \Name's inter-stage tuning pass gets the sampled pairs from the intra-tuning pass and converts the objective as \vspace{-2pt}
\begin{equation}
    \min_{\substack{1 \leq i \leq S \\ l_i, f_i, (n_i, m_i)}}
    \left\{ 
        (G-1) \cdot \max_{1 \leq i \leq S} \left\{ t_{i} \right\} + \sum_{i=1}^{S} t_{i} + \max_{1 \leq i \leq S} \left( d_i - \sum_{1 \leq j < i} t_{i}  \right) 
    \right\}
    \label{equ:inter-op-obj-with-pareto}
\end{equation}
\begin{equation}
    (t_i, d_i) = IntraStagePareto(i, l_i, (n_i, m_i))\left[ f_i \right]
\end{equation}
where $f_i$ is the sampled index from the intra-stage Pareto frontier introduced in the next section.
We directly combine the checkpointing tuning into the Pareto frontier as it also serves as a trade-off between $t_i$ and $d_i$.
Objective~\eqref{equ:inter-op-obj-with-pareto} can be reformulated into an MILP problem and solved by the off-the-shelf solver~\cite{forrest2005cbc}.

% \subsubsection{Symbolic-Analysis based Intra-stage Tuning}
\myparagraph{Intra-Stage Tuning}
As Objective~\eqref{equ: intra-tuning-high-level-objective} shows, given the stage partitioning, device assignment, gradient accumulation steps, and memory budget,
intra-stage tuning finds the best data and tensor parallelism, and memory optimization combinations to maximize the throughput and sample the Pareto frontier. % of combinations of the parallelism and memory footprint optimizations listed in Tab~\ref{tab:optim-schedule-explaination}. 
{
\begin{equation}
\begin{aligned}
    & \quad \; \min_{p, z, o} \ \alpha \cdot G \cdot t_{p, z, o} + (1-\alpha) \cdot d_{p, z, o}   \\
    & i.e. \max \left ( Mem^{(fwd)}_{peak}, Mem^{(bwd)}_{peak} \right) \leq Mem_{Budget}
    \label{equ: intra-tuning-high-level-objective}
\end{aligned}
\end{equation}
}where a series of $\alpha \in [0, 1]$ are sampled uniformly to construct a Pareto frontier efficiently. And the stable microbatch time $t$ and delta time $d$ of a certain gradient accumulation step $G$ and strategy tuple ( parallelism $p$, ZeRO config $z$, and offloading configs $o$) can be obtained from the interference model. The parallelism strategy $p$ includes $b$, $DP$, $TP$. The offloading configuration $o$ consists of $OO$, $GO$, $WO$, and $AO$.
\begin{align}
    t_{p, z, o} = \mathcal{I}\left(  c_{p, z, o}^{stable}, nccl_{p, z, o}^{stable}, d2h_{p, z, o}^{stable}, h2d_{p, z, o}^{stable}  \right) \\
    d_{p, z, o} = \mathcal{I}\left(  c_{p, z, o}^{first}, nccl_{p, z, o}^{first}, d2h_{p, z, o}^{first}, h2d_{p, z, o}^{first}  \right) - t_{p, z, 0}
\end{align}
where $\mathcal{I}$ is the interference model proposed before, $c$ means the GPU computation time, $nccl$ means the GPU-GPU communication time, $d2h$ means the device to host copy time, and $h2d$ means the host to device copy time. The superscript $stable$ indicates the time of a stable micro batch, while $first$ indicates the time of the first micro batch.

All the statistics of runtime and memories are reported by our symbolic analyzer. With the help of our symbolic-based performance analyzer, querying single datapoints is extremely fast.
Thus, to get the best strategy, we simply search in a brute-force way, which would not miss any optimization possibilities, ensuring the optimal solution.
\section{Evaluation}

We prototype \Name with $\sim$27K LoC in Python.
To support all optimizations, we have implemented it from scratch based on PyTorch~\cite{pytorch}, supporting symbolic torch tracing and execution, model automatic pipelining, overlapped offloading and ZeRO execution, and memory buffer optimizations.

We evaluate \Name on various training configurations with different hardware, models, and hyper-parameters to demonstrate its ability to effectively find the optimal combination of memory optimizations and parallelism.
Our results show that \Name constantly outperforms state-of-the-art distributed training systems. 
We use training throughput (samples per second) as our primary metric.
Since all optimizations applied by \Name are lossless, the fidelity of computation is preserved, ensuring the model convergence is not affected. 
Additionally, we provide speedup breakdown, sensitivity studies, prediction accuracies, tuning time, and case studies to explore the sources of our speedup and provide insights.

\subsection{Methodology}

\begin{table}[!t]
% \footnotesize
\scriptsize
\centering
\caption{Hardware Specifications.}
 \setlength{\tabcolsep}{2.0pt}
 \begin{tabular}{c c c c c c c c}
 \toprule
    \textbf{Platform} & \textbf{GPU} & \textbf{GPU\#} & \textbf{Mem.} & \textbf{PCIe Spec} & \textbf{NVLink} &  \textbf{Interconnect} \\
 \midrule
    GCP & L4   & [2, 4, 8, 16, 32] & 24GB & Gen3@16x & \xmark & 100Gbps \\
    AWS & A100 & [2, 4, 8, 16, 32] & 40GB & Gen4@16x & \cmark & 400Gbps \\
 \bottomrule
 \end{tabular}
\label{tab:hardware-spec}
\end{table}
\begin{table}[!t]
\scriptsize
\centering
\caption{Workload Specifications.}
 \setlength{\tabcolsep}{2.0pt}
 \begin{tabular}{c c c c c c c}
 \toprule
    \textbf{GPU} & \textbf{Models} & \textbf{Param\# (billion)} & \textbf{Global Batch Size} & \textbf{Seq Len} \\
 \midrule
    L4 & GPT, Llama, Falcon & [1.3, 2.6, 6.7, 13, 22] & [32, 64, 128, 256, 512] & 2048 \\
    A100 & GPT, Llama, Falcon & [1.3, 2.6, 6.7, 13, 22] & [32, 64, 128, 256, 512] & 4096 \\
 \bottomrule
 \end{tabular}
\label{tab:workload-spec}
\end{table}

\myparagraph{Hardware Settings}
To fully study the capabilities of \Name, we evaluate its training performance on both PCIe and NVLink systems, as they offer different combinations of hardware resources.
We conduct our major experiments on up to 32 NVIDIA L4 GPUs~\cite{nvidia-l4} and 32 NVIDIA A100 GPUs~\cite{nvidia-a100}.
Detailed hardware specifications are shown in Table~\ref{tab:hardware-spec}.
% We conduct our major experiments on up to 4 GCP G2 VMs, each equipped with 8 NVIDIA L4 24GB GPUs. The maximum network bandwidth is 100Gbps~\cite{gcp-l4-bandwidth}.
% Additionally, we evaluate \Name on up to 4 AWS EC2 p4d.24xlarge instances, each equipped with 8 NVIDIA A100-SXM4-40GB GPUs. The GPUs in these instances are connected via NVLinks which provide up to 600GB/s theoretical intra-node bandwidth. The instances are connected by 400Gbps InfiniBand with AWS Elastic Fabric Adapter (EFA) and GPUDirect RDMA enabled~\cite{aws-p4d.24xlarge}.

\myparagraph{Workloads Setting}
We selected three representative types of LLMs, GPT-3\cite{brown2020gpt3}, LLaMa~\cite{llama, llama2, llama3}, and Falcon~\cite{almazrouei2023falcon}. 
All are transformer-based models with some variation in their components. 
GPT-3 consists of typical transformer decoder layers\cite{brown2020gpt3}.
LLaMa integrates techniques like pre-RMSNorm~\cite{zhang2019rmsnorm}, gated functions~\cite{shazeer2020glu}, rotary embedding~\cite{su2024rotary}, among others, to improve performance on tasks involving long-range dependencies~\cite{llama, llama2, llama3}. 
Falcon adopts parallel attention and MLP layers inspired by GPT-J~\cite{chen2021gpt-j} and GPT-NeoX~\cite{gpt-neox-20b}, reducing the number of all-reduce operations associated with tensor parallelism from two to one per layer~\cite{almazrouei2023falcon}. Following common practice, we scale the number of GPUs and the global batch size with the size of the model.
To minimize the impact of different frameworks and kernel implementations, we set the dropout ratio to zero and disable all biases in the linear layers.
Table~\ref{tab:workload-spec} shows the workloads specifications.

\myparagraph{Baselines}
We compare \Name with three state-of-the-art deep learning distributed training systems: (1) Megatron-LM~\cite{megatron-1} (\texttt{core\_r0.4.0}), (2) DeepSpeed~\cite{deepspeed} (\texttt{v0.12.6}), and (3) Aceso~\cite{liu2024aceso}. \textbf{Megatron-LM} and \textbf{DeepSpeed} are state-of-the-art manual implementations. Since they do not support automatic tuning, to achieve the best performance, we perform a grid search over all possible optimization combinations for single-node distributed training. 
    For multi-node cases, we benchmark the best strategies that \Name finds within the same search space as Megatron-LM and DeepSpeed.  \textbf{Aceso} is the state-of-the-art automatic distributed strategy tuner with search space larger than others~\cite{chen2023slapo, sun2024adapipe}, which can automatically find the best combinations of parallelism and activation checkpointing plans. We follow its artifact to get its numbers.

In the common practice of training LLMs, FlashAttention~\cite{dao2022flashattention, dao2023flashattention2}, a vital kernel for performing the fast and memory-efficient attention mechanism, is applied by default to reduce memory usage and achieve the best performance. 
When FlashAttention is enabled, we only compare with Megatron-LM and DeepSpeed since the Aceso does not support it.
We compare all three baselines on L4 GPUs.
For A100 GPUs, we only compare \Name with state-of-the-art manual and automatic methods, Megatron-LM and Aceso as DeepSpeed generally underperforms Megatron-LM in our experiments.
% However, we only compare \Name with Megatron-LM and Aceso for multi-node distributed training on A100 GPUs due to (1) the fact that Megatron-LM and Aceso usually outperforms DeepSpeed and Alpa, respectively, (2) resource constraint of the A100 GPU cluster availability.

We attempted to compare with Alpa~\cite{zheng2022alpa}, but it fails to find any feasible solutions on L4 GPUs for our workloads. 
Our conjecture is that Alpa only considers memory usage in the Inter-Op pass by compiling the searched strategy and running it, while its memory-unaware Intra-Op pass likely causes OOM errors for all proposed strategies.
% Since its Intra-Op pass is not memory-aware, it's possible that all strategies proposed by Intra-Op pass causes OOM errors.

\begin{figure*}[t]
    \centering
    \includegraphics[width=0.92\textwidth]{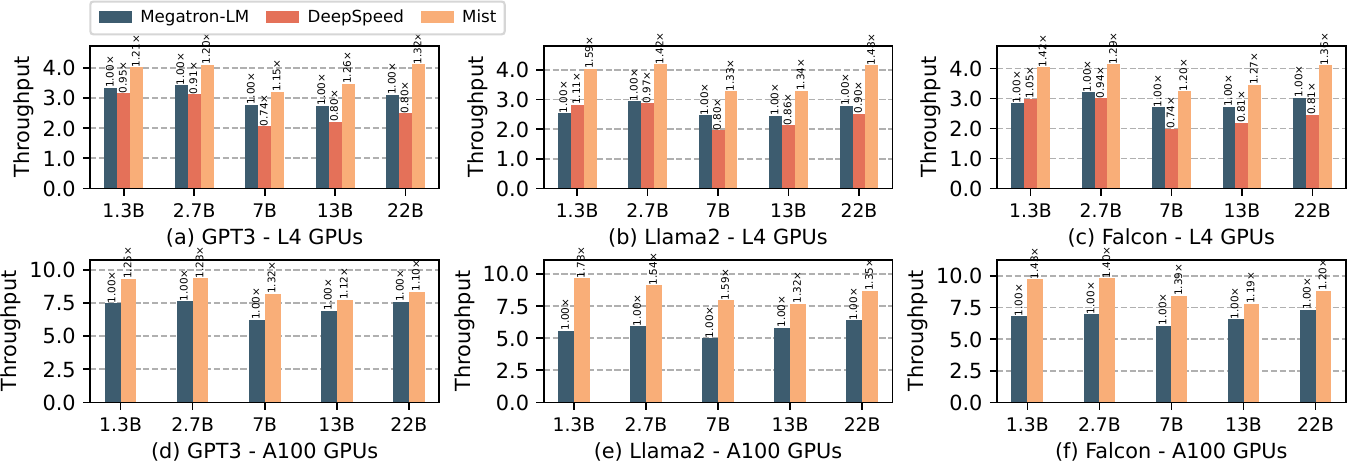}
    \caption{End-to-end training throughput (samples/sec) on L4 GPUs and A100 GPUs, with FlashAttention enabled. Sequence lengths are 2048 for L4 GPUs and 4096 for A100 GPUs. The numbers of GPUs are 2, 4, 8, 16, 32, respectively.}
    \label{fig:eval-1-main-with-FlashAttn}
\end{figure*}

\begin{figure*}[t]
    \centering
    \includegraphics[width=0.95\textwidth]{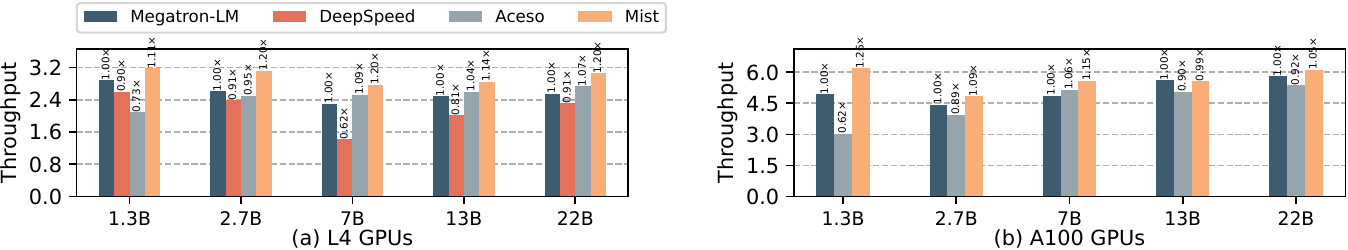}
    \caption{End-to-end training throughput (samples/sec) on L4 GPUs and A100 GPUs, without FlashAttention. Sequence lengths are 2048 for L4 GPUs and 4096 for A100 GPUs. The numbers of GPUs are 2, 4, 8, 16, 32, respectively.}
    \label{fig:eval-2-main-without-FlashAttn}
\end{figure*}

\subsection{End-to-End Training Performance}
\label{sec:e2e-perf}

\myparagraph{Speedup in Real-World Scenarios} 
Figure~\ref{fig:eval-1-main-with-FlashAttn} compares the end-to-end throughput of various distributed training frameworks in real-world scenarios where FlashAttention~\cite{dao2022flashattention, dao2023flashattention2} is enabled.
We make two key observations.
First, Megatron-LM outperforms DeepSpeed in most cases. 
This is mainly due to the fact that the parallelization plans that work in Megatron-LM cause out-of-memory issues in DeepSpeed, forcing DeepSpeed to choose sub-optimal parallelization strategies.
Second,
\Name consistently outperforms other distributed training frameworks, achieving an average speedup of 1.32$\times$ (up to 1.59$\times$) over Megatron-LM on L4 GPUs, 1.51$\times$ on average (up to 1.67$\times$) over DeepSpeed on L4 GPUs, and 1.34$\times$ on average (up to 1.72$\times$) over Megatron-LM on A100 GPUs.
Specifically for the GPT-3 model, which is the most heavily optimized model in other frameworks, \Name achieves 1.22$\times$ speedup on average (up to 1.32$\times$) on L4 GPUs, and 1.20$\times$ speedup on average (up to 1.32$\times$) on A100 GPUs, compared with Megatron-LM.
The higher speedup for LLaMa model mainly comes from the better RMSNorm kernel implementation and efficient rotary embedding implementation~\cite{dao2022flashattention}.
Overall, we conclude that \Name achieves the best performance over prior state-of-the-art manual implementations across various models and hardware.

\myparagraph{Speedup compared with more baselines}
Figure~\ref{fig:eval-2-main-without-FlashAttn} compares the throughput of the GPT-3 model with both manual and automatic parallelization frameworks without FlashAttention. 
\Name still consistently outperforms or is equal to all prior distributed training frameworks.
\Name achieves an average of 1.14$\times$ speedup (up to 1.26$\times$ speedup) compared to Megatron-LM and an average of 1.27$\times$ speedup (up to 2.04$\times$ speedup) compared to Aceso.
When training GPT-3 13B on 16 A100 GPUs, \Name does not achieve better performance but still gets almost the same results, because the naive strategy happens to achieve the best trade-off among all resources.
We also find that Aceso does not consistently outperform Megatron-LM even though it has larger search space due to fine-grained activation checkpointing tuning. 
The root cause is that Aceso does not include sharded data parallelism in the search space and miss several essential opportunities for communication-computation overlapping.

\myparagraph{Discussion on the hardware}
As shown in Figures~\ref{fig:eval-1-main-with-FlashAttn} and ~\ref{fig:eval-2-main-without-FlashAttn}, \Name exhibits higher speedup on L4 GPUs than that on A100 GPUs, with following reasons.
Large-scale distributed training tasks on L4 GPUs are often limited by smaller memory capacity and the restricted intra-node and inter-node bandwidth. 
In this scenario, \Name plays a crucial role in striking the best trade-off among various resources to enhance the resource utilization.
On the other hand, training tasks on A100 GPUs benefit from larger memory capacity and faster intra-node NVLink and inter-node InfiniBand connections, resulting in much higher resource utilization that approaches the physical limits.
This leaves less room for improvement.

\subsection{Speedup Breakdown}
\begin{figure}[t]
    \centering
    \includegraphics[width=0.98\columnwidth]{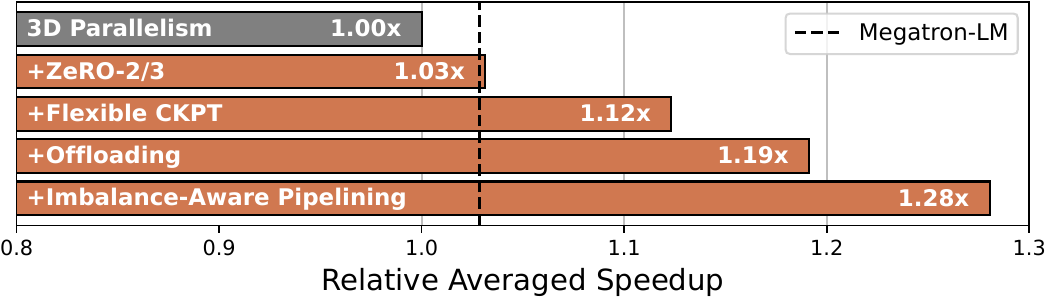}
    \caption{Relative averaged speedup of tuning over different search spaces for GPT model on 8, 16, and 32 L4 GPUs.}
    \label{fig:eval-3-ablation}
\end{figure}

To understand how each key ideas of \Name contributes to the final performance, we evaluate it by incrementally enlarging the search space proposed by \Name in Figure~\ref{fig:eval-3-ablation}. We normalized the throughput by the baseline search space of Megatron-LM.
Three key conclusions are drawn:
First, \Name's advantage is not from the better implementation; with the same search space as Megatron-LM, \Name is slightly slower due to implementation overhead supporting other optimizations.
Second, activation checkpointing tuning provides a 1.12$\times$ speedup on average, with offloading adding an extra 7\% speedup, showing their abilities of striking better trade-offs among resources.
Third, inter-microbatch imbalance-awareness offers an extra 9\% speedup upon all prior speedups, as it provides accurate runtime predictions for pipeline parallelism. % when sharded data parallelism and optimizer offloading are involved.
In summary, all optimizations included in \Name are  crucial for improve the system performance.

One key insight we observe is that much of the speedup comes from reducing activation checkpointing and eliminating pipeline bubbles. 
However, naively disabling checkpointing often leads to OOM errors, and tuning it (as done in Aceso) only partially solves the issue. 
Further improvements from increasing ZeRO levels or increasing offloading are essential, as they help to further reduce recomputation.
As long as these overheads can be overlapped or amortized across multiple microbatches, performance significantly improves.

Additionally, speedups may vary depending on hardware resources and workload intensity.
On A100 GPUs with moderate workloads, most speedups come from activation checkpointing tuning.
However, when memory pressure is high, combining it with offloading becomes important.
For instance, training a 40B GPT-3 model on 32 A100 GPUs, \Name is expected to get 1.10$\times$ speedup compared to 1.04$\times$ with only activation checkpointing tuning.

\begin{figure}[t]
    \centering
    \includegraphics[width=1.0\columnwidth]{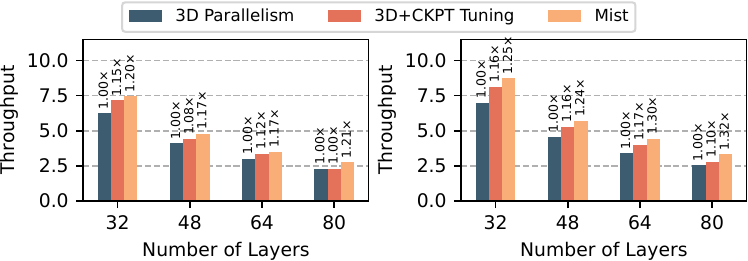}
    \caption{Performance of GPT-3 with different number of layers on 32 L4 GPUs. Left: without FlashAttention; Right: with FlashAttention.}
    \label{fig:eval-4-different-scale}
\end{figure}

\subsection{Sensitivity Study}

To comprehensively understand the robustness of \Name, we evaluate its performance with different model scales and different global batch sizes.

\myparagraph{Robustness over Different Model Scales}
As depicted in Figure~\ref{fig:eval-4-different-scale}, \Name consistently outperforms the baseline search spaces by up to 1.32$\times$ higher throughput, particularly at 80 layers, regardless of whether FlashAttention is enabled.
Activation checkpointing tuning is particularly effective for smaller model sizes.
However, as model size increases, the speedup from checkpointing alone decreases.
With the entire search space enabled, \Name maintains substantial speedups across different model sizes.

\myparagraph{Robustness over Different Global Batch Sizes}
\begin{figure}[t]
    \centering
    \includegraphics[width=0.929\columnwidth]{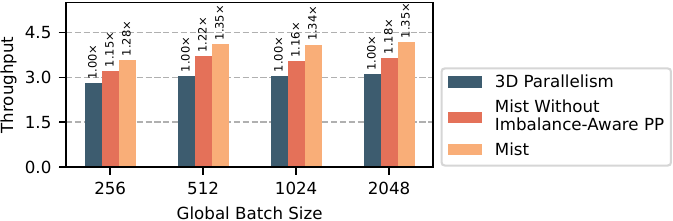}
    \caption{Performance of GPT-3 22B with different global batch sizes on 32 L4 GPUs.}
    \label{fig:eval-5-different-gbs}
\end{figure}
As shown in Figure~\ref{fig:eval-5-different-gbs}, \Name always achieves the best performance compared to the baseline search space across different global batch sizes.
Notably, Imbalance-Aware Inter-Stage Tuning provides an extra 1.13$\times$ speedup on average.
One concern is that with larger global batch sizes and potentially more microbatches, the benefit of imbalance-aware inter-stage tuning might diminish, as treating all microbatches as the same might seem sufficient.
However, the sub-optimal strategy produced by inaccurate predictions lead to a significant performance gap due to the larger gradient accumulation steps while \Name's inter-microbatch awareness avoids it.

\begin{figure}[t]
    \centering
    \includegraphics[width=0.85\columnwidth]{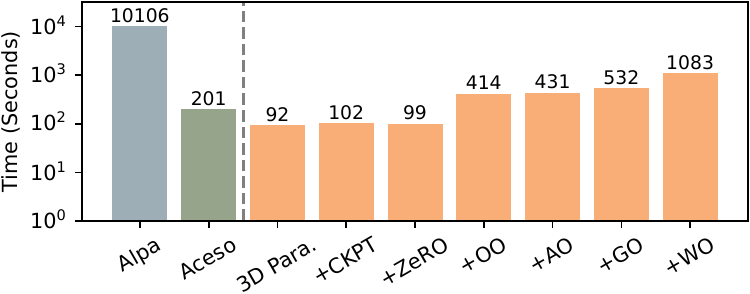}
    \caption{Tuning Time of the GPT-3 22B model on 32 GPUs. Orange bars are \Name with different optimizations applied.}
    \label{fig:eval-tuning-time}
\end{figure}

\subsection{Tuning Time Comparison}
\label{sec:tuning-time-comparison}
As Figure~\ref{fig:eval-tuning-time} shows, to understand the tuning efficiency of \Name, we evaluate the tuning time by enabling optimizations one by one and compare it with the tuning time of Alpa~\cite{zheng2022alpa} and Aceso~\cite{liu2024aceso}.
For Alpa, we choose 6 data points because it doesn't automatically tune the gradient accumulation steps and layer grouping size.
We make three key observations:
First, \Name helps to reduce the tuning time a lot compared to Alpa.
Second, when \Name is configured to use a similar search space as Aceso, which is branded as an efficient distributed training searching system, \Name can be faster.
Third, even when \Name enables more optimization and greatly increases the search space, the tuning time remains reasonable compared to the significantly longer training time.
Moreover, since searching over different gradient accumulation steps is independent, \Name's tuning can be parallelized across different CPU cores or machines to make it faster.

\subsection{Accuracy of Symbolic Shape Analysis System}
\label{sec:accurate-of-symbolic-analysis}
The effectiveness of \Name's tuning system relies heavily on the performance prediction accuracy. 
Therefore, we sample different strategies and benchmark the accuracy of the predicted runtime and memory usage compared with the actual ones.
\Name consistently demonstrates a high prediction accuracy for both runtime and memory usage. 
The averaged runtime error ratio is 1.79\%, and average memory footprint error ratio is 2.10\%.
For runtime, our analysis system focuses more on the magnitude comparison to determine the best strategy. As a result, some minor times, such as the optimizer step time, are not included. 
To better understand runtime accuracy, we shift the predicted runtime so that the mean values of predicted and actual runtime match.

\section{Other Related Work}

% \myparagraph{Parallel Training Frameworks}
% Parallel training frameworks such as Megatron and Deepspeed \cite{megatron-1, megatron-2, megatron-3, deepspeed, zero, zero-offload, zero-infinity}, although they incorporate many optimization techniques supported by \Name, they lack the automatic tuning of \Name, and thus rely on manual user intervention. This is also the case with GSPMD \cite{gspmd} that requires the user to manually add annotations in their code. There are a few prior works that support automatic parallelization tuning, such as Slapo, TensorOpt, Alpa, Piper, and Tofu \cite{chen2023slapo, cai2020tensoropt, zheng2022alpa, wang2019tofu, tarnawski2021piper}, although they lack the full range of co-optimization provided by \Name, which as we motivate in \cref{sec: } is needed to significantly improve system performance in distributed training.

\myparagraph{DNN Performance Modeling}
The closest works to ours in performance modeling are DistSim, Proteus, and dPRO \cite{distsim, duan2023proteus, hu2022dpro}, which use simulation-based approaches to predict model performance. \Name differs in that we consider an expanded search space including all memory optimizations in addition to parallelism strategies. This requires additional sophisticated modeling with respect to overlap and interference. Moreover, \Name employs a symbolic modeling first approach that is well-suited for the efficiency that the expanded search space demands. DistIR \cite{santhanam2021distir} is a work that also employs an analytical-based approach, but employs a cost-model predictor of operator latencies. Habitat \cite{geoffrey2021habitat} profiles operators in tandem with methods to extrapolate performance to other GPUs for non-distributed scenarios. 

\myparagraph{Symbolic Analysis}
Symbolic analysis examines program behavior through abstract representations of variables and computations rather than concrete values, enabling systematic exploration of optimization spaces. Foundational applications in compilers leveraged symbolic techniques for dependency analysis and loop transformations via constraint solving~\cite{fahringer1997symbolic,cadar2008klee}, while analog circuit design adopted symbolic methods for parameter space exploration~\cite{gielen1994symbolic-circuit-design, floberg2012symbolic-circuit-design}. 
Unlike prior works targeting code-level optimizations or hardware verification, \Name adapts symbolic analysis in the area of performance estimation for deep learning models to efficiently explore the joint space of parallelism strategies and memory reduction techniques in distributed training.

\myparagraph{Acceleration Techniques}
Techniques like tensor compilation and kernel optimizations (e.g., TVM, Hidet, FlashAttention) are largely different than \Name, since they mainly focus on lower-level, static graph optimizations from the graph to hardware level \cite{tvm, hidet, xla, ansel2024pytorch, he2023transcending, dao2022flashattention, dao2023flashattention2, lin2023efficient}. %There is some overlap with methods that, for example enhance checkpointing \cite{checkmate} or Slapo's "model definition to execution" \cite{chen2023slapo}. 
Additionally, approaches such as gradient compression, quantization, and sparsity, and automatic mixed precision \cite{bai2021super, wang2023cupcake, actnn, chen2022butterfly, mixed} are also orthogonal to \Name: they could be integrated into \Name's schedule template as additional optimization techniques that can improve system performance and memory efficiency. Some of these may also be potentially lossy optimizations, i.e., downgrading the accuracy of models, whereas \Name exclusively targets system-level improvements without compromising training accuracy.
\section{Discussion}

\myparagraph{Integration of Other Techniques}
\Name is extensible to additional operators, parallelism strategies, and optimizations. New acceleration or memory optimization techniques like quantization~\cite{mixed, peng2023fp8} or compression~\cite{actnn,liu2022gact} are typically implemented using native torch operators or customized kernels. Native torch operators (including communication operators) can be straightforwardly supported by \Name, and customized CUDA kernels can be easily incorporated by registering them in the symbolic analysis system, as demonstrated with FlashAttention~\cite{dao2022flashattention}. Therefore, these optimizations are directly reflected in the traced computational graph produced by \Name and analyzed by \Name, thus they will be configured intelligently to achieve high performance.

\myparagraph{Future Work}
\Name relies on a symbolic computational graph, making it less suited for highly dynamic workloads where a fixed computation graph is difficult to obtain. However, for workloads like Mixture of Experts (MoE) with expert parallelism~\cite{rajbhandari2022deepspeed-moe}, where computation patterns are largely predictable, data-dependent routing can be handled through multiple simulations to obtain an average performance estimate.
Another limitation is that, while \Name supports fine-grained overlap of multiple operations, ensuring correctness remains challenging due to potential data races and value inconsistencies. Future work should focus on developing automated mechanisms to manage overlap and prevent execution conflicts.
Additionally, although \Name can analyze arbitrary computational graphs, its efficient tuning algorithm assumes identical layers. Extending it to optimize models with heterogeneous architectures is an important direction.
\section{Conclusion}
We propose \Name, a memory, overlap, and imbalance aware method that enables efficient LLM training by co-optimizing all memory footprint reduction techniques and parallelism strategies in a comprehensive manner. \Name contributes a fine-grained overlap-centric schedule template, an symbolic-based efficient analysis system, and an imbalance-aware hierarchical auto-tuner to allow efficient optimization in a large search space over optimizations with complex interactions. As a result, \Name achieves 1.27$\times$ (up to 2.04$\times$) over state-of-the-art distributed training systems such as Aceso. We hope that \Name will be able to help democratize LLM training for machine learning researchers and practitioners alike.

%%
%% The acknowledgments section is defined using the "acks" environment
%% (and NOT an unnumbered section). This ensures the proper
%% identification of the section in the article metadata, and the
%% consistent spelling of the heading.
\begin{acks}
We sincerely thank our shepherd, Zhaoguo Wang, and the anonymous reviewers for their valuable feedback.
We also appreciate members of the EcoSystem Research Laboratory at the University of Toronto for their discussions and suggestions, with special thanks to Anand Jayarajan, Xin Li, Wei Zhao, Yaoyao Ding, and Jiacheng Yang for their contributions.
The authors with the University of Toronto are supported by Vector Institute Research grants, the Canada Foundation for Innovation JELF grant, NSERC Discovery grant, AWS Machine Learning Research Award (MLRA), Facebook Faculty Research Award, Google Scholar Research Award, and VMware Early Career Faculty Grant.

\end{acks}

%%
%% The next two lines define the bibliography style to be used, and
%% the bibliography file.
\bibliographystyle{ACM-Reference-Format}
\bibliography{references}

%%
%% If your work has an appendix, this is the place to put it.
\appendix

%%%%%%%%%%%%%%%%%%%%%%%%%%%%%%%%%%%%%%%%%%%%%%%%%%%%
% Artifact Appendix Template for EuroSys'25 AE
%
% this document has a maximum length of 2 pages.
%%%%%%%%%%%%%%%%%%%%%%%%%%%%%%%%%%%%%%%%%%%%%%%%%%%%

\newpage
\section{Artifact Appendix}

%%%%%%%%%%%%%%%%%%%%%%%%%%%%%%%%%%%%%%%%%%%%%%%%%%%%%%%%%%%%%%%%%%%%%
\subsection{Abstract}

\Name is an automatic distributed training configuration optimizer designed to tune the optimal configuration for the combination of parallelism strategies and memory footprint reduction techniques. 
This artifact includes the source code of the \Name prototype along with instructions for evaluating its functionality and reproducing key results.

%%%%%%%%%%%%%%%%%%%%%%%%%%%%%%%%%%%%%%%%%%%%%%%%%%%%%%%%%%%%%%%%%%%%%
\subsection{Description \& Requirements}

\subsubsection{How to access}

% The \Name system is available at the following Github repository: \href{https://github.com/dazz993/mist}{https://github.com/dazz993/mist} and ZENODO: DOI: 10.5281/zenodo.14873554.
The \Name system is available at the following repositories: 
\href{https://github.com/dazz993/mist}{https://github.com/dazz993/mist} and \href{https://doi.org/10.5281/zenodo.14873554}{Zenodo (DOI: 10.5281/zenodo.14873554)}

\subsubsection{Hardware dependencies}

Experiments are conducted on up to four GCP L4 machines, each equipped with 8$\times$ NVIDIA L4 GPUs, and up to four AWS p4d.24xlarge machines, each with 8$\times$NVIDIA A100 40GB GPUs. For artifact evaluation, an ideal testbed consists of a machine with 8 $\times$ GPUs, each with approximately 24GB of memory, as we can provide configurations suitable for direct evaluation on such hardware. Unless explicitly stated otherwise, the following evaluation workflow assumes this testbed. A general evaluation methodology for other GPUs and multi-node evaluation is also provided in our repository.

\subsubsection{Software dependencies}

We provide a Docker image with NVIDIA GPU support for this artifact. The software environment includes CUDA 12.1, PyTorch v2.1.1, Megatron-LM (git-hash 38879f8), DeepSpeed (v0.12.6), and NCCL v2.18.6.

\subsubsection{Benchmarks} 
None.

%%%%%%%%%%%%%%%%%%%%%%%%%%%%%%%%%%%%%%%%%%%%%%%%%%%%%%%%%%%%%%%%%%%%%
\subsection{Set-up}

To install the artifact, users should clone the repository and build the Docker image. For users with GPUs other than L4 GPUs (sm\_89), the  environment variable \texttt{TORCH\_CUDA\_ARCH\_LIST} in the Dockerfile may require modification.
\vspace{3pt}
\begin{lstlisting}[language=bash]
git clone https://github.com/Dazz993/Mist.git
cd Mist
docker build -t mist -f Dockerfile .
\end{lstlisting}

\noindent Run the Docker container, mounting the \Name repository home to \texttt{/workspace}.
\vspace{3pt}
\begin{lstlisting}[language=bash]
docker run --gpus all -it --rm --privileged \
  --ipc=host --shm-size=20G --ulimit memlock=-1 \
  --name "mist" -v $(pwd):/workspace/ mist
\end{lstlisting}

\noindent [Optional] To obtain stable results, especially on L4 machines, fix the GPU frequency accordingly:
\vspace{3pt}
\begin{lstlisting}[language=bash]
nvidia-smi -ac 6251,1050
\end{lstlisting}

%%%%%%%%%%%%%%%%%%%%%%%%%%%%%%%%%%%%%%%%%%%%%%%%%%%%%%%%%%%%%%%%%%%%%
\subsection{Evaluation workflow.}

We provide scripts for reproducing single-node results. Multi-node experiments require large-scale clusters and additional setup and execution time, so end-to-end scripts are not provided for artifact evaluation. However, instructions are available in GitHub repository README. \textbf{We recommend users to follow the README as it provides extra explanations.}

\subsubsection{Major Claims}

\begin{itemize}
    \item \textit{(C1)}: Mist achieves an average of 1.28$\times$ (up to 1.73$\times$) speedup compared to state-of-the-art manual system Megatron-LM. See Section~\ref{sec:e2e-perf} and Figure~\ref{fig:eval-1-main-with-FlashAttn} and ~\ref{fig:eval-2-main-without-FlashAttn}. This claim is validated by E2.
    \item \textit{(C2)}: Mist demonstrates efficient tuning speed even with a large search space. See Section~\ref{fig:eval-tuning-time} and Figure~\ref{fig:eval-tuning-time}. This claim is validated by E3.
    % \item \textit{(C3)}: Mist provides accurate performance and memory usage predictions. The averaged runtime error ratio is 1.79\% and the average memory footprint error ratio is 2.1\%. See Section~\ref{sec:accurate-of-symbolic-analysis}.
\end{itemize}

\subsubsection{Experiments}

\begin{itemize}[leftmargin=11pt, itemsep=0pt, topsep=3pt]
    \item E1: Kick-the-Tries [10 human-minutes]. This experiment evaluates the functionalities of \Name on Large Language Model analysis, execution, and distributed training optimization. Detailed explanations and expected results are also provided in the repository README.

    \textit{[Execution]} Given a YAML configuration for running the GPT-3 1.3B model on two GPUs: \texttt{test-small-base}:

\begin{itemize}[leftmargin=11pt, itemsep=0pt, topsep=2pt]
    \item Run model performance analysis
\end{itemize}
\begin{lstlisting}[language=bash]
cd /workspace/benchmark/mist/analysis/
python run.py --config-name test-small-base
\end{lstlisting}

\begin{itemize}[leftmargin=11pt, itemsep=0pt, topsep=2pt]
    \item Execute the model distributed training
\end{itemize}
\begin{lstlisting}[language=bash]
cd /workspace/benchmark/mist/exec/
torchrun --nproc-per-node 2 \
    benchmark_one_case.py \
    --config-name test-small-base
\end{lstlisting}

\begin{itemize}[leftmargin=11pt, itemsep=0pt, topsep=2pt]
    \item Run model tuning (the hyperparameters in this configuration file are specifically tuned for GCP L4 GPUs).
\end{itemize}
\begin{lstlisting}[language=bash,basicstyle=\ttfamily\tiny]
cd /workspace/benchmark/mist/tune/
python tune_one_case.py --config-name test-small-base \
    +output_path=/workspace/benchmark/mist/tune/results/test-small-mist
\end{lstlisting}
\quad Then execute the optimized configuration:
\begin{lstlisting}[language=bash,basicstyle=\ttfamily\scriptsize]
cd /workspace/benchmark/mist/exec/
torchrun --nproc-per-node 2 \
    benchmark_one_case.py \
    --config-path /workspace/benchmark/mist/tune/results/ \
    --config-name test-small-mist
\end{lstlisting}

\textit{[Results]} The executed commands output the analysis results, execution time, and memory usage for the base configuration, as well as the execution time and memory usage for the optimized configurations.

\item E2: Run Single-Node Performance Evaluation [4 compute-hours]. This experiment evaluates the performance of \Name on a single node for GPT and LLaMA models.

For the L4 machine, we provide pre-tuned configurations that enable a quick assessment of Mist’s speedup compared to baseline models. Additionally, we provide a general process for evaluating on a new cluster. For further details, refer to the GitHub repository README.

\textit{[Execution]}

\begin{itemize}[leftmargin=11pt, itemsep=0pt, topsep=2pt]
    \item Evaluate Mist performance. Results are summarized in /workspace/benchmark/mist/tuned\_configs/l4-24gb/ gpt/summary.json and corresponding LLaMA folder.
\end{itemize}
\begin{lstlisting}[language=bash]
cd /workspace/benchmark/mist/tuned_configs/
bash run_single_node.sh
\end{lstlisting}

\begin{itemize}[leftmargin=11pt, itemsep=0pt, topsep=2pt]
    \item Evaluate Megatron-LM performance. Results are under /workspace/benchmark/megatron/results.
\end{itemize}
\begin{lstlisting}[language=bash,basicstyle=\ttfamily\footnotesize]
cd /workspace/benchmark/megatron/
bash scripts/tops/l4/gpt2/1_8xl4_node_1_pcie.sh
bash scripts/tops/l4/llama/1_8xl4_node_1_pcie.sh
\end{lstlisting}

\begin{itemize}[leftmargin=11pt, itemsep=0pt, topsep=2pt]
    \item Evaluate DeepSpeed performance. Results are under /workspace/benchmark/deepspeed/results.
\end{itemize}
\begin{lstlisting}[language=bash,basicstyle=\ttfamily\footnotesize]
cd /workspace/benchmark/deepspeed/
bash scripts/tops/l4/gpt2/1_8xl4_node_1_pcie.sh
bash scripts/tops/l4/llama/1_8xl4_node_1_pcie.sh
\end{lstlisting}

\textit{[Results]} We provide a python file to collect the results for easy comparison. 
\begin{lstlisting}[language=bash,basicstyle=\ttfamily\footnotesize]
cd /workspace/benchmark/
python scripts/collect_single_node_results_v1.py
\end{lstlisting}

It provides tables with absolute throughput values or relative throughput improvements. Here we provide an example:
\begin{lstlisting}[language=bash,basicstyle=\ttfamily\scriptsize, backgroundcolor=\color{white}]
+-----------------------+-------------+-------------+
| SpeedUp                  | SpeedUp vs   | SpeedUp vs    |
|                          | Megatron     | DeepSpeed     |
+=======================+=============+=============+
| gpt2-1.3b-flash_False | 1.175X          | 1.418X        |
+-----------------------+-------------+-------------+
| gpt2-2.7b-flash_False | 1.141X          | 1.384X        |
+-----------------------+-------------+-------------+
| gpt2-7.0b-flash_False | 1.222X          | 2.053X        |
+-----------------------+-------------+-------------+
\end{lstlisting}

\item E3: Benchmarking Tuning Time [0.75 compute hours]. This experiment reproduces the tuning time results for Mist, as shown in Figure~\ref{fig:eval-tuning-time}.

We demonstrate how tuning time varies as the search space expands incrementally. To evaluate this, we run a GPT-22B model on a 4 × 8 GPU setup. Beyond examining tuning time, this experiment also provides insights into the performance of large-scale distributed training, as we can see the performance improvements when more optimizations are applied.

\textit{[Execution]}
\vspace{2pt}

\begin{lstlisting}[language=bash,basicstyle=\ttfamily\footnotesize]
cd /workspace/benchmark/mist/benchmark-tuning-time
python run.py --model=gpt2/22b -n 4 -m 8
\end{lstlisting}

The results are shown in /workspace/benchmark/mist/ benchmark-tuning-time/results/.../summary.json.

\end{itemize}

%%%%%%%%%%%%%%%%%%%%%%%%%%%%%%%%%%%%%%%%%%%%%%%%%%%%%%%%%%%%%%%%%%%%%
\subsection{Notes on Reusability}
\label{sec:reuse}
Mist provides a symbolic execution system that represents all tensors with symbolic dimensions. Additionally, it supports tracing, which generates a corresponding symbolic computational graph. This feature can serve as an educational tool, helping users understand shape propagation and how each input dimension is utilized. Furthermore, it serves as a basis for exploring performance estimation in future research.

\end{document}